\documentclass[aps,twocolumn,superscriptaddress,PRD]{revtex4}
\usepackage{amsmath}
\usepackage{epsfig}
\usepackage{multirow}
\usepackage{slashed}
\usepackage{amsmath}
\usepackage{color}
\usepackage{float}
\usepackage{graphicx}
\usepackage[colorlinks,
            linkcolor=blue,
            anchorcolor=blue,
            citecolor=blue
            ]{hyperref}

\def\bea#1\eea{\begin{align}#1\end{align}}

\newcommand{\bef}{\begin{figure}[!htp]}
\newcommand{\eef}{\end{figure}}

\begin{document}
\title{Imaging nuclear modifications on parton distributions with triple-differential\\
dijet cross sections in proton-nucleus collisions}

\author{Shuwan Shen}
\affiliation{Key Laboratory of Quark $\&$ Lepton Physics (MOE) and Institute of Particle Physics, Central China Normal University, Wuhan 430079, China}

\author{Peng Ru}
\email{p.ru@m.scnu.edu.cn}
\affiliation{Guangdong Provincial Key Laboratory of Nuclear
Science, Institute of Quantum Matter, South China Normal University, Guangzhou 510006, China}
\affiliation{Guangdong-Hong Kong Joint Laboratory of Quantum
Matter, Southern Nuclear Science Computing Center, South China Normal University, Guangzhou 510006, China}

\author{Ben-Wei Zhang}
\email{bwzhang@mail.ccnu.edu.cn}
\affiliation{Key Laboratory of Quark $\&$ Lepton Physics (MOE) and Institute of Particle Physics, Central China Normal University, Wuhan 430079, China}
\affiliation{Guangdong Provincial Key Laboratory of Nuclear
Science, Institute of Quantum Matter, South China Normal University, Guangzhou 510006, China}

\date{\today}

\begin{abstract}
Dijet production in proton-nucleus~($p$A) collisions at the LHC provides invaluable
information on the underlying parton distributions in nuclei, especially the gluon
distributions. Triple-differential dijet cross sections enable a well-controlled kinematic
scan~(over momentum fraction $x$ and probing scale $Q^2$) of the nuclear parton distribution
functions~(nPDFs), i.e., $f^\textrm{A}_i(x,Q^2)$. In this work, we study several types
of triple-differential cross sections for dijet production in proton-proton~($pp$) and
proton-lead~($p$Pb) collisions at the LHC, to next-to-leading order within the framework
of perturbative quantum chromodynamics~(pQCD).
Four sets of nPDF parametrizations, EPPS16, nCTEQ15, TUJU19, and nIMParton16 are employed
in the calculations for $p$Pb collisions. We show that the observable nuclear modification
factor $R_{p\textrm{Pb}}$ of such cross sections can serve as a nice
image of the nuclear modifications on parton distributions, quantified by the ratio
$r^{\textrm{A}}_i(x,Q^2)\!=\!f^\textrm{A,proton}_i(x,Q^2)/f^\textrm{proton}_i(x,Q^2)$.
Considerable differences among the $R_{p\textrm{Pb}}$ predicted by the four nPDF sets
can be observed and intuitively understood. Future measurements of such observables are expected
 to not only constrain the nPDF parametrizations, but also help confirm various nuclear
 effects, e.g., shadowing, anti-shadowing, EMC, and Fermi motion in different regions of $x$
 and their variations with probing scale $Q^2$.
\end{abstract}

\maketitle

%%%%%%%%%%%%%%%%%%%%%%%%%%%%%%%%%%%%%%%%%%%%%%%%
\section{Introduction}
Nuclear parton distribution functions~(nPDFs) are indispensable non-perturbative inputs in
the study of various hard scattering processes in high-energy nuclear collisions~\cite{
Paukkunen:2018kmm,Kovarik:2019xvh}, including lepton-nucleus, hadron-nucleus and nucleus-nucleus
collisions. Due to the additional dynamics that bind nucleons together, the partonic structure
of a large nucleus can obviously deviate from that of the free nucleons~\cite{Arneodo:1992wf,
Armesto:2006ph,Cloet:2012td,CLAS:2019vsb,Kulagin:2004ie,Ru:2016wfx,Wang:2020uhj}. To exactly
determine this deviation will not only help understand the interactions among bound nucleons,
but also provide a baseline to precisely study other physics, such as the final-state jet
quenching phenomena in hot and dense quark-gluon plasma in relativistic nucleus-nucleus
collisions~\cite{Cao:2020wlm,Gyulassy:2003mc}.

The deviations of the nuclear parton distributions from the free-proton ones are
usually quantified with the ratios $r^{\textrm{A}}_i(x,Q^2)\!=\!f^\textrm{A,proton}_i(x,Q^2)/f^\textrm{proton}_i(x,Q^2)$, which
are generally functions of the nuclear mass number $A$, parton flavor $i$, momentum fraction $x$
and the resolution scale $Q^2$. Currently, there exist many parameterized nPDF sets determined
through the global analysis of world experimental data~\cite{deFlorian:2011fp,Eskola:2009uj,Eskola:2016oht,
Eskola:2021mjl,Hirai:2007sx,Khanpour:2020zyu,Kovarik:2015cma,Kusina:2020dki,AbdulKhalek:2020yuc,Wang:2016mzo,
Walt:2019slu,Segarra:2020gtj},
based on the factorization in perturbation
theory of quantum chromodynamics~(QCD). However, differences among the factors $r^A_i(x,Q^2)$
given by these nPDF sets can be observed, and their uncertainties are still considerable~\cite{Paukkunen:2018kmm}.

Before the running of the Large Hadron Collider~(LHC), the main data sources
of the global extraction are from lepton deeply inelastic scattering~(DIS) off nucleus and
Drell-Yan~(DY) process in hadron-nucleus collisions. In both of them a detailed scan of nPDFs over
$x$ and $Q^2$ can be achieved by controlling the final-state kinematic variables.
For example, in DIS the conventional two variables are Bjorken $x_B$ and photon virtuality,
and in DY the invariant mass and rapidity of dilepton are typically chosen~\cite{Ellis:QCD}.
Nevertheless, both DIS and DY are largely sensitive to the quark distributions, but provide
less direct constraints on gluon distribution. Thus, the uncertainty of the nuclear correction
$r^A_i(x,Q^2)$ on gluon distribution is usually greater than that of the quark distribution~\cite{Eskola:2009uj}.
Although some early global analyses have taken into account the pion data from the Relativistic Heavy Ion
Collider~(RHIC), which are more sensitive to the nuclear gluon distribution, the unavoidable
fragmentation functions may introduce an additional theoretical uncertainty in the determination of nPDFs~\cite{CMS:2018jpl}.

The wealthy data from the LHC have significantly promoted the extraction of nPDFs. For example,
the production of weak boson through DY process provides a new insight into nuclear quark
distributions at a high resolution scale~\cite{Paukkunen:2010qg,Ru:2014yma,Ru:2016wfx,Ru:2015pfa,Albacete:2013ei}.
In particular, powerful constraints on nuclear gluon distribution are given by the productions of dijets~\cite{CMS:2018jpl,
Eskola:2019dui} and heavy-quark mesons~($J/\Psi$, $D^0$, etc.)~\cite{Kusina:2017gkz,Kusina:2020dki}
in proton-nucleus~($p$A) collisions, with which both shadowing and anti-shadowing effects on
gluons have been more consolidated~\cite{Eskola:2019dui,Kusina:2017gkz,Kusina:2020dki}.

Dijet production in $p$A collisions at the LHC has rich yields and has been well studied within
perturbative QCD~\cite{Ellis:1992en,Ellis:1994dg,Nagy:2001fj,Nagy:2003tz,Gehrmann-DeRidder:2019ibf,Kunszt:1992tn,Gao:2012he}.
Moreover, the theoretical calculations don't rely on fragmentation functions.
Thus, it is an important probe of both the quark and gluon distributions in nuclei~\cite{Eskola:2013aya,He:2011sg}.
However, since the dynamic channels for dijet production are more complicated than those in DIS and DY,
a detailed scan of the nPDFs over $x$ and $Q^2$ may not be easily achieved by controlling only two
kinematic variables of dijet. In contrast, a better correspondence to $x$ and $Q^2$ can be established
by simultaneously fixing three dijet variables, which has motivated the idea to study the triple-differential
dijet cross sections~\cite{Ellis:1994dg,CMS:2017jfq,Gehrmann-DeRidder:2019ibf} instead of
the double-differential ones. So far, this idea has been applied in $pp$ collisions
to study the proton PDFs~\cite{Ellis:1994dg,CMS:2017jfq,Gehrmann-DeRidder:2019ibf}.
As will be discussed in this paper, the triple-differential dijet cross sections have more
advantages in $p$A collisions for unveiling the nuclear corrections of PDFs.

In this work, we will study several types of triple-differential dijet cross sections in both $pp$ and $p$A
collisions to next-to-leading order~(NLO) within pQCD. Four sets of nPDFs, EPPS16~\cite{Eskola:2016oht},
nCTEQ15~\cite{Kovarik:2015cma}, TUJU19~\cite{Walt:2019slu}, and nIMParton16~\cite{Wang:2016mzo}
are used in the calculations of $p$A.
We show that the nuclear modification factors $R_{pA}$ of the triple-differential cross sections
can well disclose the $x$ and $Q^2$ dependence of the nPDF factors $r^A_i(x,Q^2)$ of quark and
gluon distributions, and can even serve as a nice image of the $r^A_i(x,Q^2)$ from small to large values of $x$.
This in turn provides an intuitive way to understand the observed differences among the $R_{pA}$
predicted by various nPDFs.

The rest of this paper is organized as follows. In Sec.~\ref{sec:baseline} we review the dijet
production in both $pp$ and $p$A collisions and the theoretical framework used in this study.
In the discussion about $pp$~(\ref{sec:pp}) we focus on the dijet kinematics and the necessity
to introduce the triple-differential cross section. In the subsection about $p$A~(\ref{sec:CNM})
we pay attention to the initial-state cold nuclear matter effects related to the nPDFs.
In Sec.~\ref{sec:triple} we study three types of triple-differential cross sections and the
corresponding nuclear modification factors $R_{pA}$. We establish links to compare the observable
$R_{pA}$ with the nPDF factors $r^A_i(x,Q^2)$. Possibilities to reveal
the variation of $r^A_i(x,Q^2)$ with scale by using $R_{pA}$ in different $p_T$ regions are also discussed.
We give a summary and discussion in Sec.~\ref{sec:summary}.

%%%%%%%%%%%%%%%%%%%%%%%%%%%%%%%%%%%%%%%%%%%
\section{Dijet production in $pp$ and $p$A collisions}
\label{sec:baseline}
\subsection{Dijet in $pp$ collisions and kinematics}
\label{sec:pp}
At partonic level, the production of dijet in $pp$ collisions is related to the processes of
$N$-parton~(quarks or gluons, $N\geq2$) production initiated by partons $a$ and $b$ from the two colliding protons.
In perturbative QCD, the cross section for the production of $N$ partons in $pp$ collisions can be generally expressed as
the convolution of the parton distribution functions $f_i(x,Q^2)$ and the hard $2\!\rightarrow\!N$ scattering cross
section $d\hat{\sigma}^{[2\rightarrow N]}_{ab}$~\cite{Gao:2012he}
\bea
\frac{d\sigma}{d\Phi_N}=\sum_{a,b}\int_0^1 \!\!\!dx_a\!\int_0^1 \!\!\!dx_b
\,f_a(x_a,\mu^2_f)\,f_b(x_b,\mu^2_f)\cr
\times\frac{\,d\hat{\sigma}^{[2\rightarrow N]}_{ab}(x_a,x_b,\mu_r,\mu_f)\,}{d\Phi_N},
\label{eq:partonic}
\eea
where $\Phi_N\equiv\Phi_N(p_1,\ldots,p_N)$ represents the phase space of the final-state $N$ partons,
$x_a$~($x_b$) is the momentum fraction carried by the incoming parton from the forward~(backward)-going proton,
and $\mu_f$~($\mu_r$) is the factorization~(renormalization) scale. The right-hand side of Eq.~(\ref{eq:partonic}) is summed over
the flavors of partons $a$ and $b$, essentially including quark-quark, quark-gluon and gluon-gluon initial states.

At NLO in perturbative calculation, where only $2\rightarrow2$ and $2\rightarrow3$ partonic
processes are involved, a dijet observable, such as an $m$-fold differential
cross section ${d\sigma}/{\prod_{j=1}^{m} dv_j}$ defined with the dijet kinematic variables
$V^{(m)}=\{v_1,\ldots,v_m\}$, can be calculated with~\cite{Gao:2012he}
\bea
\frac{d\sigma}{\prod_{j=1}^{m} dv_j}\!=\!\int d\Phi_2(p_1,p_2)\,\frac{d\sigma}{d\Phi_2}\,S_2(p_1,p_2)
\,\,\,\,\,\,\,\,\,\,\,\,\,\,\,\,\,\,\,\,\,\cr
+\!\int d\Phi_3(p_1,p_2,p_3)\,\frac{d\sigma}{d\Phi_3}\,S_2(p_1,p_2,p_3).
\label{eq:NLO}
\eea
Here all the constraints imposed on final state, including jet algorithm, are embodied in
functions $S_N(p_1,\ldots,p_N)$, which should be properly defined to satisfy the infrared safe conditions~\cite{Gao:2012he}.

Equations (\ref{eq:partonic}-\ref{eq:NLO}) link the measurable dijet cross section
to the not directly measurable parton distribution functions $f_i(x,Q^2)$, within the framework of factorization.
Conventionally, to nicely disclose the $x$ and $Q^2$ dependence of the PDFs at an observable level,
one can establish a one-to-one correspondence between the values of $V^{(m)}$ and those of $x$ and $Q^2$
in the leading-order~(LO) approximation, by properly choosing the dijet variables $V^{(m)}$.

For instance, one applicable variable set to define a double-differential cross section may be
$V^{(2)}=\{M_{{\!J\!J}},y_{\textrm{dijet}}\}$~\cite{Eskola:2013aya}, with the squared invariant mass
$M^2_{{\!J\!J}}\equiv(p_1+p_2)^2$ and rapidity $y_{\textrm{dijet}}$ of the jet pair.
Note that a similar choice is usually used in DY measurements.
With this choice, one can connect $V^{(m)}$ with $x_{a(b)}$ through the LO relation
\bea
x_{a(b)}=\frac{M_{{\!J\!J}}}{\sqrt{s}}e^{\pm y_{\textrm{dijet}}},
\label{eq:my}
\eea
which enables a kinematic scan of $x_{a(b)}$. Besides, one physical scale $M_{{\!J\!J}}$ is also well controlled.
Not like the DY process, where the typical hard scale is the invariant mass of the lepton pair,
there is an additional physical scale in dijet production, i.e., the jet transverse momentum~($p_{T1}\!=\!p_{T2}$ at LO),
which is not yet fully controlled with the above choice of $V^{(m)}$.

As a matter of fact, the number of independent kinematic variables of the dijet final state is 3
at LO~\cite{Ellis:1994dg}. Therefore, by defining a triple-differential cross section with
$V^{(3)}=\{v_1,v_2,v_3\}$, one may simultaneously resolve $x_{a(b)}$ and the scales at which
the parton distributions are probed. One straightforward example is to set
$V^{(3)}=\{M_{{\!J\!J}},y_{\textrm{dijet}},p_{T,avg}\}$, with $p_{T,\textrm{avg}}\!=\!(p_{T1}\!+\!p_{T2})/2$.
On the other hand, the high statistics at the LHC are sufficient for a precise measurement
of the more-differential observable~\cite{CMS:2017jfq}.

In general, the established LO correspondence can be more or less broken once the higher-order corrections are considered.
As a result, for a realistic observable, it is inevitable that the sub-processes associated with various
values of $x$, $Q^2$ and $i$~(flavor) jointly contribute to the measurement at a single set of $V^{(m)}$.
A demonstration in a kinematic perspective is presented as follows.

For a process of $N$-parton production, one can infer the initial-state momentum fraction
$x_a$~($x_b$) by summing the forward~(backward) components of the light-cone momenta over all the $N$
final-state partons as~\cite{Ellis:1994dg}
\bea
x_{a(b)}=\sum_{n=1}^N \frac{p_n^{\pm}}{\sqrt{s}}=\sum_{n=1}^N \frac{E_{Tn}}{\sqrt{s}}e^{\pm y_n},
\label{eq:inferx}
\eea
with the light-cone component $p_n^{\pm}\equiv p^0_n\pm p^3_n=E_{Tn}e^{\pm y_n}$, transverse energy
$E_{Tn}=(m^2+p_{Tn}^2)^{1/2}$ and rapidity $y_n$ of parton $n$.
However, in an inclusive measurement of dijet, one can not fully deduce the values of $x_a$ and $x_b$,
since the final-state particles that lie outside the dijet cones are not restricted, which could happen
at NLO and beyond.
These issues also underlie the fact that the global QCD analysis for the PDFs
is necessary and complicated.
%These issues also underlie the fact that the global QCD analysis is a complicated and
%irreplaceable approach to accessing the PDFs.

Nonetheless, since the LO processes may give a dominant contribution, one can still expect
that the triple-differential dijet cross sections have a nice resolution power for the $x$
and $Q^2$ dependence of PDFs, which is beneficial to the global analysis~\cite{CMS:2017jfq}.
%Besides, as will be discussed in Sec.~\ref{sec:triple}, triple-differential dijet cross
%section will have more advantages for the study of the nuclear modifications in $p$A collisions.

\begin{figure}[t]
\hspace{-0.5cm}\includegraphics[width=3.3in]{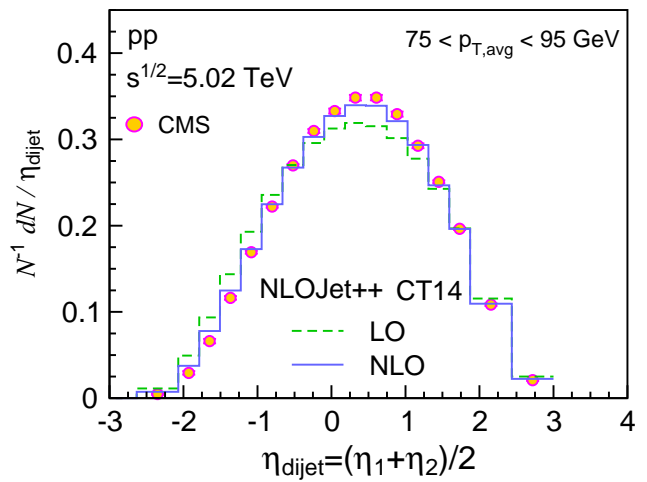}
\caption{Normalized dijet pseudo-rapidity distribution in $pp$ collisions at $\sqrt{s}=5.02$~TeV.
Here $\eta_{\textrm{dijet}}$ is defined in laboratory frame in $p$Pb collisions, thus is related to that in
center of mass frame as $\eta_{\textrm{dijet}}=\eta^{\textrm{cm}}_{\textrm{dijet}}+0.465$~\cite{CMS:2018jpl}.
Theoretical results are calculated at LO~(dashed) and NLO~(solid), with CT14 LO and NLO proton PDFs, respectively.
Circle represents CMS data~(vertical uncertainty bar is small)~\cite{CMS:2018jpl}.
Jets are found by anti-$k_T$ algorithm with cone size $R=0.3$ and with rapidity cut $|y_{\textrm{lab}}|<3$.
Averaged transverse momentum of dijet is restricted with $75<p_{T,\textrm{avg}}<95$~GeV.
Relative azimuthal angle of jet pair is restricted with $|\Delta\phi_{12}|>2\pi/3$.
Cuts imposed on transverse momenta of leading and sub-leading jets are $p_{T1}>30$~GeV and $p_{T2}>20$~GeV, respectively~\cite{CMS:2018jpl}.
}
\label{fig:eta_cms_pp}
\end{figure}

In this work, we will calculate the double- and triple-differential dijet cross sections to NLO in perturbative QCD by using the code NLOJet++~\cite{Nagy:2001fj,Nagy:2003tz}.
To test the numerical calculations, we first show in Fig.~\ref{fig:eta_cms_pp} the results for
dijet pseudo-rapidity~[$\eta_{\textrm{dijet}}\!=\!(\eta_1\!+\!\eta_2)/2$] distribution in $pp$ collisions, which has been measured by the CMS collaboration~\cite{CMS:2018jpl}.
We find that the CMS data can be well described by the NLO calculation, in which the CT14 proton PDFs~\cite{Dulat:2015mca} are used and
the factorization and renormalization scales are taken to be $\mu_0=\mu_f=\mu_r=p_{T,\textrm{avg}}$.

Since this $\eta_{\textrm{dijet}}$ distribution is measured with the averaged transverse momentum
of dijet $p_{T,\textrm{avg}}$ being restricted,
the normalized distribution $N^{-1}dN/d\eta_{\textrm{dijet}}$ shown in Fig.~\ref{fig:eta_cms_pp} actually corresponds to the
double-differential cross section $\sigma^{-1}d^2\sigma/d\eta_{\textrm{dijet}}dp_{T,\textrm{avg}}$ with
$V^{(2)}=\{\eta_{\textrm{dijet}},p_{T,\textrm{avg}}\}$.
Please note that, even at LO, neither the dijet invariant mass $M_{{\!J\!J}}$ nor $x_{a(b)}$
are uniquely determined with a certain set of $\{\eta_{\textrm{dijet}},p_{T,\textrm{avg}}\}$,
while the ratio $x_a/x_b=e^{2\eta_{\textrm{dijet}}}$ is fixed.
Overall, the averaged value of $x_{a(b)}$ will increase~(decrease) with an increasing
$\eta_{\textrm{dijet}}$~\cite{CMS:2018jpl}.
We leave a further discussion in Sec.~\ref{sec:triple}.

\subsection{Dijet in $p$A collisions and cold nuclear matter effects}
\label{sec:CNM}
In $p$A collisions, the differences between nuclear PDFs and free-nucleon PDFs
may result in nuclear modifications on dijet cross section relative to that in $pp$ collisions,
which are also referred to as the initial-state cold nuclear matter~(CNM) effects~\cite{Albacete:2017qng}.
The nuclear correction on parton distribution is usually quantified with the ratio
of the PDF in the bound nuclear proton to that in free proton as
\bea
r^A_i(x,Q^2)=\frac{f^{A,p}_i(x,Q^2)}{f^{p}_i(x,Q^2)}.
\label{eq:ri}
\eea
Generally, the nuclear correction factors $r^A_i(x,Q^2)$ are determined by the additional non-perturbative
dynamics with the presence of nuclear environment, and they depend on the nuclear mass number $A$, parton flavor $i$, momentum fraction $x$
and the resolution scale $Q^2$. Phenomenologically, the nuclear effects, i.e., $r^A_i(x,Q^2)\ne 1$, are conventionally
classified, from small to large values of $x$, as the shadowing, anti-shadowing, EMC, and Fermi motion~\cite{Arneodo:1992wf,Armesto:2006ph}.

\begin{figure}[t]
\hspace{-0.5cm}\includegraphics[width=3.0in]{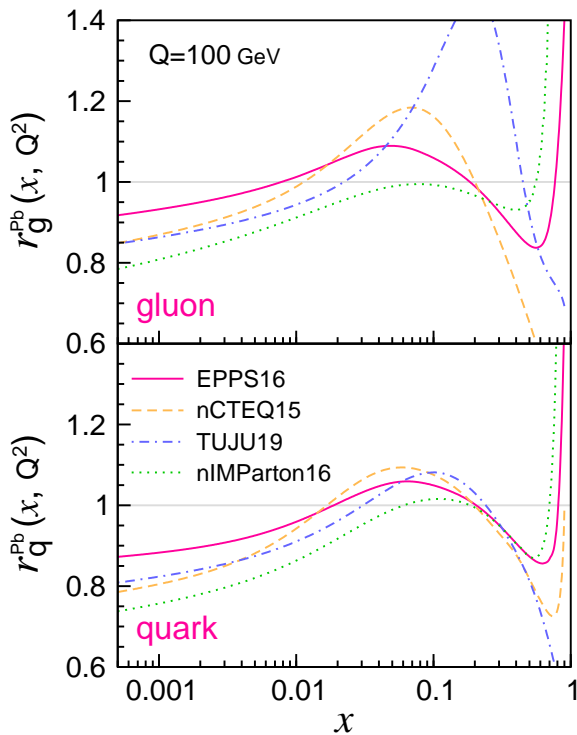}
\caption{Nuclear correction factors $r^{\textrm{Pb}}_i(x,Q^2)$ for gluon~(top panel) and quark~(bottom panel) distributions
in lead nucleus from four nPDF sets, EPPS16, nCTEQ15, TUJU19, and nIMParton16 at $Q=100$~GeV. }
\label{fig:ratio_nPDF}
\end{figure}

In this work, to include the initial-state CNM effects in $p$A collisions, we use four parametrization
sets of nuclear PDFs, i.e., EPPS16~\cite{Eskola:2016oht}, nCTEQ15~\cite{Kovarik:2015cma}, TUJU19~\cite{Walt:2019slu},
and nIMParton16~\cite{Wang:2016mzo} in the calculations.
In Fig.~\ref{fig:ratio_nPDF}, we plot the factors $r^A_i(x,Q^2)$ for gluon and quark distributions in lead~($^{208}$Pb) nucleus
from the four nPDF sets, at the scale $Q=100$~GeV. Considerable differences among these nPDF sets
can be observed, especially for gluon distribution.
For some specific values of the variable $x$, some nPDFs give enhancements and others can give suppressions, which means the ranges
of each kind of nuclear effect are not yet clear.
These issues also motivate our study for a more delicate kinematic scan of the nPDFs in $p$A collisions.
In this work, we will focus on how these differences are reflected in dijet observables. We hope our
study is helpful for the future measurements to clarify the ranges of these nuclear effects.

In our calculations for dijet production, to unify the baseline $pp$ results, we have used the factors $r^{A}_i(x,Q^2)$ from
the four nPDF sets on top of the CT14 proton PDFs for $p$A collisions. We note TUJU and nIMP nPDFs are
extracted with their own proton PDFs other than CT14, and we have neglected this difference since we only focus on
the nuclear modifications in this work. In addition, because nIMP is obtained in a LO QCD analysis~\cite{Wang:2016mzo},
we only do the LO calculation when using nIMP. The other three nPDF sets are used in NLO calculations.
In this study, we have only considered the nuclear corrections on the leading-twist PDFs and not focused on the
possible higher-twist nuclear effects~\cite{Kang:2013raa,Ru:2019qvz}. Besides, since there has been no clear
experimental evidence of the possible final-state jet quenching effects in $p$A collisions~\cite{Xie:2020zdb,Huss:2020whe},
we have only considered the initial-state CNM effects.

\begin{figure*}[!t]
\centering
\hspace{-0.5cm}\includegraphics[width=4.5in]{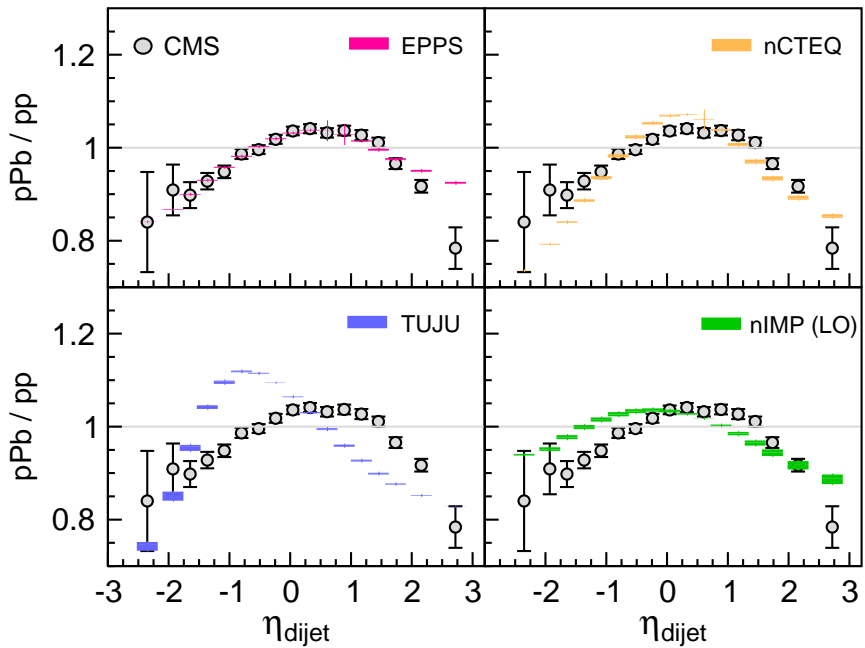}
\caption{Ratio of normalized dijet $\eta_{\textrm{dijet}}$ distribution in $p$Pb and $pp$ collisions, corresponding to
Fig.~\ref{fig:eta_cms_pp}. Circle with vertical bar represents CMS data with uncertainty~\cite{CMS:2018jpl}.
Colored lines represent theoretical predictions with four nPDFs sets, i.e., EPPS16, nCTEQ15, TUJU19 and nIMParton16,
with bands corresponding to variations with factorization/renormalization scales ($\mu_0/2, \mu_0, 2\mu_0$) in perturbative calculations.
Results with nIMP are calculated at LO, while results with other three nPDF sets are calculated at NLO. }
\label{fig:eta_cms_rpa_sep}
\end{figure*}

As an example, we calculate the normalized $\eta_{\textrm{dijet}}$ distribution in proton-lead~($p$Pb) collisions, corresponding to
that in $pp$ collisions shown in Fig.~\ref{fig:eta_cms_pp}, and plot the ratio of the results in $p$Pb and $pp$ collisions
in Fig.~\ref{fig:eta_cms_rpa_sep}, where the CMS data~\cite{CMS:2018jpl} are also shown for comparison.
Differences among the predictions with four nPDF sets can be seen,
and since the CMS measurement is very precise, the experimental uncertainties can be even smaller than the differences among various
predictions in a wide range of $\eta_{\textrm{dijet}}$.
The results with EPPS show a nice overall agreement with the data, except for that in forward region related to
the shadowing at small $x$. This is consistent with the results in Ref.~\cite{CMS:2018jpl}. We also note that the EPPS reweighted
with the CMS dijet data indeed suggests a stronger shadowing suppression in gluon distribution~\cite{Eskola:2019dui}.
The results with nCTEQ can also well describe the data. However, the predictions with both TUJU and nIMP show
deviations from the data to some extents. For example, in the region around $\eta_{\textrm{dijet}}\sim -1$, TUJU
predicts a strong enhancement, which is not observed in the data. This enhancement is related to the
strong gluon anti-shadowing effect as seen in Fig.~\ref{fig:ratio_nPDF}~($x\sim0.2$).
We note that TUJU also provides NNLO nPDFs, in which a weaker anti-shadowing effect is suggested~\cite{Walt:2019slu}.

It should be mentioned that the aim of this study is not to rule out any parameterized nPDFs, but to provide a more intuitive
way for studying the differences among various nPDF sets and to facilitate more effective measurements in future.
Thus, in the calculations, we only use the central values of the nPDFs without considering their uncertainties.
In Fig.~\ref{fig:eta_cms_rpa_sep}~(and also in following figures), the bands for the theoretical results correspond to the variations
with factorization/renormalization scales ($\mu_0/2, \mu_0, 2\mu_0$) in the perturbative calculations.

Due to the high accuracy, the CMS data have already provided a powerful constraint on the nPDFs, especially on
nuclear gluon distribution~\cite{Eskola:2019dui}. However, the initial-state parton momentum fraction $x$
is not fully controlled at LO as discussed in Sec.~\ref{sec:pp}, which results in the mixing of different
nuclear effects in the observable. Next, we discuss the possibility to make further improvement
on the observable level, i.e., to reduce the extent of the mixing by using triple-differential dijet cross section.

~

~

~

\section{Triple-differential dijet cross section and nuclear modifications in proton-lead collisions}
\label{sec:triple}
As is discussed in Sec.~\ref{sec:pp}, triple-differential dijet cross section provides a more direct resolution of the
$x$ and $Q^2$ dependence of the PDFs $f_i(x,Q^2)$. This idea has been applied to $pp$ collisions to study the proton PDFs~\cite{Ellis:1994dg,CMS:2017jfq,
Gehrmann-DeRidder:2019ibf}. However, in $pp$ collisions, even a triple-differential
cross section can still involve two different momentum fractions, $x_a$ and $x_b$, carried by the two initial partons~\cite{Gehrmann-DeRidder:2019ibf}.
Besides, to make a more adequate comparison with a measured cross section, the NNLO corrections may be needed
in a theoretical prediction~\cite{Gehrmann-DeRidder:2019ibf}. In contrast, in the study of nuclear modifications in $p$A collisions,
these problems can be avoided to a large extent, and the triple-differential cross section
has more advantages. On one hand, the parton incoming from the single nucleus target only involves one momentum fraction variable,
whereas the parton from proton can be viewed as a probe. On the other hand, if we focus on the nuclear correction
factors $r^A_i(x,Q^2)$ instead of the nPDFs themselves, both the theoretical and experimental
uncertainties can be reduced effectively.

Concretely, one can study the conventional nuclear modification factor for such cross sections
defined as the ratio of the cross sections in $p$A and $pp$ collisions normalized by the nuclear mass number $A$~\cite{LHCb:2017ygo}
\bea
R_{pA}(v_1,v_2,v_3)=\frac{1}{A}\frac{d\sigma^{pA}/dv_1dv_2dv_3}{d\sigma^{pp}/dv_1dv_2dv_3}.
\label{eq:rpa}
\eea
This $R_{pA}$ can be schematically expressed in LO approximation as~\cite{Ellis:1994dg}
\bea
R_{pA}(v_1,\!v_2,\!v_3)\!\approx\!\frac{\sum_{a,b}f_a^p(x_a,\mu^2)f_b^A(x_{b},\mu^2)H_{ab}(v_1,\!v_2,\!v_3)}
{\sum_{a,b}f_a^p(x_a,\mu^2)f_b^{p}(x_b,\mu^2)H_{ab}(v_1,\!v_2,\!v_3)},
\label{eq:rpa-pdf}
\eea
where $f_b^A$ is the averaged per nucleon PDF in the nucleus, and $H_{ab}$ represents the perturbatively calculable hard function.
Apparently, this $R_{pA}$ provides an intuitive insight into the nuclear correction factor $r^A_i(x,Q^2)$ defined in Eq.~(\ref{eq:ri}).
Please note that, for a given set of $\{v_1,v_2,v_3\}$, the values of variables $x_{a(b)}$, $M_{{\!J\!J}}$ and $p_{T,\textrm{avg}}$ are
uniquely determined at LO.

Since the $R_{pA}$ is defined as a ratio in Eq.~(\ref{eq:rpa}), the total uncertainties
in both theoretical and experimental sides can be reduced significantly.
In theoretical calculations, the uncertainties~(e.g.~from proton PDFs and high-order corrections)
of the numerator and denominator may cancel each other to some extents.
Similarly, systematical uncertainties may be reduced effectively in experimental measurements.
Next, we will study the $R_{p\textrm{Pb}}$ for several types of
triple-differential dijet cross sections in $p$Pb collisions at the LHC.

\subsection{Nuclear modifications on $d^3\sigma/dp_{T,avg}dy_bdy^*$}
\label{sec:ptavg}
\begin{figure}[t]
\hspace{-0.5cm}\includegraphics[width=2.8in]{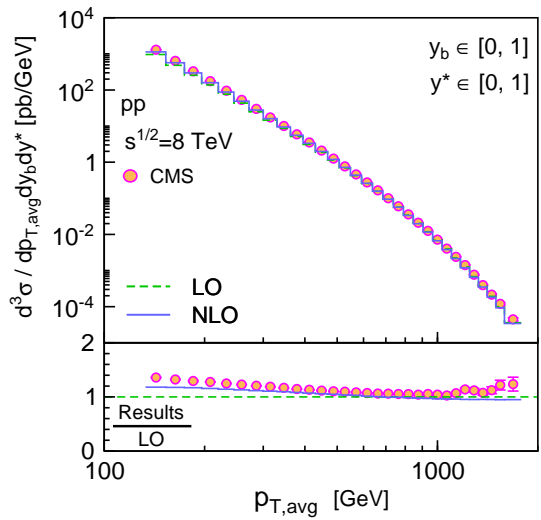}
\caption{Triple-differential cross section $d^3\sigma/dp_{T,\textrm{avg}}dy_bdy^*$ in $pp$ collisions
at $\sqrt{s}=8$~TeV with $y_b\in[0,1]$ and $y^*\in[0,1]$.
Theoretical results at LO~(dashed) and NLO~(solid) are compared to CMS data represented with
circles~(vertical uncertainty bars are small)~\cite{CMS:2017jfq}.
Jets are found by anti-$k_T$ algorithm with cone size $R=0.7$ and with rapidity cut $|y|<5$.
Additional cuts imposed on transverse momentum and rapidity of leading jet are $p_{T1}>50$~GeV
 and $|y_1|<3$, respectively~\cite{CMS:2017jfq}.
Ratios between shown results and LO ones are plotted at bottom as a reference.}
\label{fig:ptavg_8_cms_pp}
\end{figure}

We first study the nuclear modification on triple-differential dijet cross section with
$\{v_1,v_2,v_3\}=\{p_{T,\textrm{avg}},y_b,y^*\}$ that has been measured in $pp$ collisions by CMS collaboration~\cite{CMS:2017jfq}.
Here $y_b$ and $y^*$ are defined with the rapidity of the two jets as $y_b=|y_1+y_2|/2$ and $y^*=|y_1-y_2|/2$.
For convenience, we write the LO kinematic relations as~\cite{Gehrmann-DeRidder:2019ibf}
\bea
x_{a}=\frac{M_{{\!J\!J}}}{\sqrt{s}} e^{\pm y_b},~~~~x_{b}=\frac{M_{{\!J\!J}}}{\sqrt{s}} e^{\mp y_b},\cr
\textrm{with}~~~ M_{{\!J\!J}}=2p_{T,\textrm{avg}}\cosh(y^*),~~
\label{eq:ptybys_LO}
\eea
where the symbol $\pm(\mp)$ is owing to the definition of $y_b$ as an absolute value.
In Fig.~\ref{fig:ptavg_8_cms_pp}, we plot the results for $d^3\sigma/dp_{T,\textrm{avg}}dy_bdy^*$
against $p_{T,\textrm{avg}}$ in $pp$ collisions at $\sqrt{s}=8$~TeV, with $y_b\in[0,1]$ and $y^*\in[0,1]$.
The CMS data can be well described by the perturbative calculations, in which both the factorization and renormalization
scales are taken to be $\mu_0=p_{T1}e^{0.3y^*}$~\cite{CMS:2017jfq}, which is a compromise between
$p_T$ and $M_{{\!J\!J}}/2$ as first investigated in Ref.~\cite{Ellis:1992en}. The NNLO corrections have been found to
give $\sim10\%$ or more enhancements in Ref.~\cite{Gehrmann-DeRidder:2019ibf}.
Next, we apply this cross section in $p$Pb collisions in the same kinematic regions.
\begin{figure}[t]
\vspace{-0.6cm}\hspace{0.2cm}\includegraphics[width=2.8in]{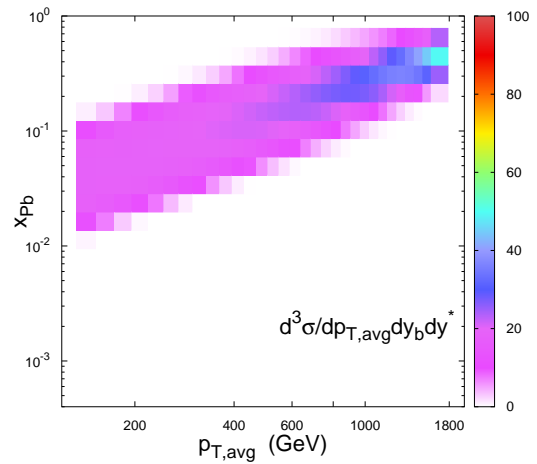}
\caption{Correlations between each $p_{T,\textrm{avg}}$ bin and momentum fraction $x_{\textrm{Pb}}$ carried by
initial nuclear parton for triple-differential dijet cross section $d^3\sigma/dp_{T,\textrm{avg}}dy_bdy^*$.
For each $p_{T,\textrm{avg}}$ bin, total contributions from various values of $x_{\textrm{Pb}}$ are normalized to be unity~($100\%$).
Values of correlations are represented with rainbow colors~(0-100\%). They are calculated by counting cross sections
in $p_{T,\textrm{avg}}$ and $x_b$ bins in $pp$ collisions~(without nuclear effects) at LO.}
\label{fig:ptavg_cr_lo_grid}
\end{figure}

\begin{figure}[!h]
\hspace{-0.8cm}\includegraphics[width=2.8in]{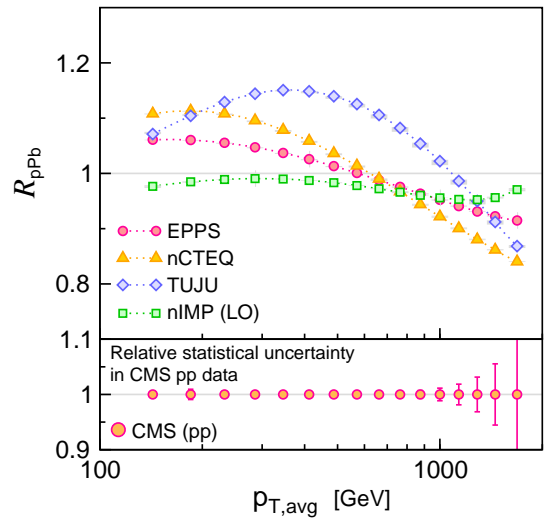}
\caption{Top panel: nuclear modification factor $R_{p\textrm{Pb}}$ corresponding to result in Fig.~\ref{fig:ptavg_8_cms_pp}.
Bottom panel: relative statistical uncertainties in CMS $pp$ data in Fig.~\ref{fig:ptavg_8_cms_pp}.
$R_{p\textrm{Pb}}$ predicted by nIMP is calculated at LO, and those by other three nPDF sets are calculated at NLO.
Grey bands~(very thin) on $R_{p\textrm{Pb}}$ results correspond to variations with factorization/renormalization scales ($\mu_0/2, \mu_0, 2\mu_0$).
Note that vertical axes in both two panels are set in same scale to make a direct comparison.}
\label{fig:ptavg_8_cms_rpa}
\end{figure}

To visualize the capability of this $d^3\sigma/dp_{T,\textrm{avg}}dy_bdy^*$ to resolve the momentum fraction $x_{\textrm{Pb}}$
of the initial parton from lead nucleus, we illustrate in Fig.~\ref{fig:ptavg_cr_lo_grid} the correlations between
each $p_{T,\textrm{avg}}$ bin and $x_{\textrm{Pb}}$, calculated at LO and represented with the rainbow colors.
It is noted that, for a certain $p_{T,\textrm{avg}}$ bin, the $x_{\textrm{Pb}}$ can spread over an order of
magnitude due to the non-vanishing bin sizes of $\Delta p_{T,\textrm{avg}}$, $\Delta y_b$ and $\Delta y^*$ in the measurement.
Nevertheless, an overall linear relation is observed as expected from Eq.~(\ref{eq:ptybys_LO}).

The nuclear modification factors $R_{p\textrm{Pb}}$ for $p$Pb collisions
are calculated with three nPDFs sets~(EPPS, nCTEQ and TUJU) at NLO and with nIMP at LO,
and are shown in the top panel of Fig.~\ref{fig:ptavg_8_cms_rpa}.
Differences among the four predictions as well as the $p_{T,\textrm{avg}}$ dependence of the results can be seen.
Since the underlying $x_{\textrm{Pb}}$ is around $0.1$ as seen in Fig.~\ref{fig:ptavg_cr_lo_grid}, which is near the anti-shadowing region,
we can see the enhancements predicted by several nPDF sets. Suppressions at large $p_{T,\textrm{avg}}$ which correspond to the EMC region
can also be observed. Currently, there has been no measurement of the triple-differential cross section in $p$Pb collisions.
To roughly estimate the constraining power of the future measurement, we show the relative statistical uncertainties in the
CMS $pp$ data~\cite{CMS:2017jfq} in the bottom panel of Fig.~\ref{fig:ptavg_8_cms_rpa} for comparison.
We find in a wide range of $p_{T,\textrm{avg}}$, the
relative uncertainties are much smaller than the differences among the nuclear modifications by various nPDFs, indicating
a strong constraint on nPDFs. We also see that the uncertainties in high-$p_{T,\textrm{avg}}$ region are too large to
distinguish the EMC effects, related to the rapidly decreasing dijet yields with increasing $p_{T,\textrm{avg}}$.

\begin{figure}[t]
\hspace{-0.3cm}\includegraphics[width=3.2in]{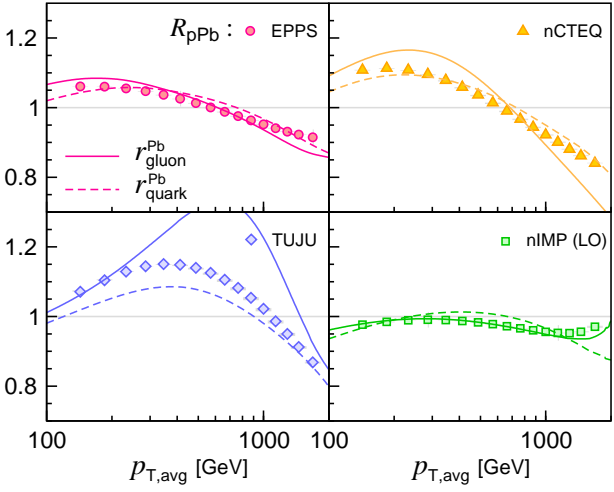}
\caption{Comparison between $R_{p\textrm{Pb}}$~(symbols) predicted by each nPDF set and its underlying nuclear corrections
$r_i^{\textrm{Pb}}(x_{\textrm{Pb}}, Q^2)$ on quark~(dashed) and gluon~(solid) distributions, corresponding to Fig.~\ref{fig:ptavg_8_cms_rpa}.
Both $R_{p\textrm{Pb}}$ and $r_i^{\textrm{Pb}}$ are shown as functions of $p_{T,\textrm{avg}}$.
For plotting $r_i^{\textrm{Pb}}$, LO relation between $x_{\textrm{Pb}}$
and $p_{T,\textrm{avg}}$ in Eq.~(\ref{eq:ptybys_LO}) is used with $y_b=0$ and $y^*=0$ being set,
and scale in $r_i^{\textrm{Pb}}(x_{\textrm{Pb}}, Q^2)$ is taken to be $Q=p_{T,\textrm{avg}}e^{0.3y^*}$.}
\label{fig:ptavg_rpa_rnpdf}
\end{figure}

Since we focus on the advantage of the triple-differential cross section on disclosing the underlying nuclear correction factor $r_i^{A}(x,Q^2)$~[Eq.~(\ref{eq:rpa-pdf})], here we make a comparison between the observable
$R_{p\textrm{Pb}}$ and the nPDF factor $r_i^{\textrm{Pb}}$ in Fig.~\ref{fig:ptavg_rpa_rnpdf}.
Please note that the factors $r_i^{\textrm{Pb}}$ for gluon and quark distributions are plotted against $p_{T,\textrm{avg}}$
instead of $x_{\textrm{Pb}}$. To make this variable substitution, we have used the LO relation in Eq.~(\ref{eq:ptybys_LO}) as
$x_{\textrm{Pb}}=2p_{T,\textrm{avg}}/\sqrt{s}$ with $y_b$ and $y^*$ taken to be 0. Meanwhile, the scale $Q$ in
$r_i^{\textrm{Pb}}(x,Q^2)$ is taken to be $Q=p_{T,\textrm{avg}}e^{0.3y^*}$ when it is plotted.
We find that the $R_{p\textrm{Pb}}$ predicted by various nPDF sets are comparable with their own $r_i^{\textrm{Pb}}$,
even though three sets of them~(EPPS, nCTEQ and TUJU) are calculated at NLO where the linear relation may be somewhat broken.
The comparison in Fig.~\ref{fig:ptavg_rpa_rnpdf} provides an intuitive way to understand the different
$R_{p\textrm{Pb}}$ given by different nPDF sets. For example, we can see that the strong anti-shadowing in gluon
distribution of TUJU significantly enhances the predicted $R_{p\textrm{Pb}}$.

The correspondence between $p_{T,\textrm{avg}}$ and $x_{\textrm{Pb}}$ plays an important role in this comparison.
However, in a single $p_{T,\textrm{avg}}$ bin as in the CMS measurement, the mixing of the contributions
from various values of $x_{\textrm{Pb}}$ is too strong to allow a more delicate kinematic scan, as shown in Fig.~\ref{fig:ptavg_cr_lo_grid}.
This mixing could be lessened by narrowing the bin sizes in the measurement of $d\sigma/dp_{T,\textrm{avg}}dy_bdy^*$,
at the cost of statistical accuracy. On the other hand, there are different ways to define the triple-differential
cross sections, which will be discussed in the following subsections.

\subsection{Nuclear modifications on $d^3\sigma/dX_BdX_Ady^*$}
\label{sec:XBXA}
\begin{figure}[b]
\hspace{-0.5cm}\includegraphics[width=3.5in]{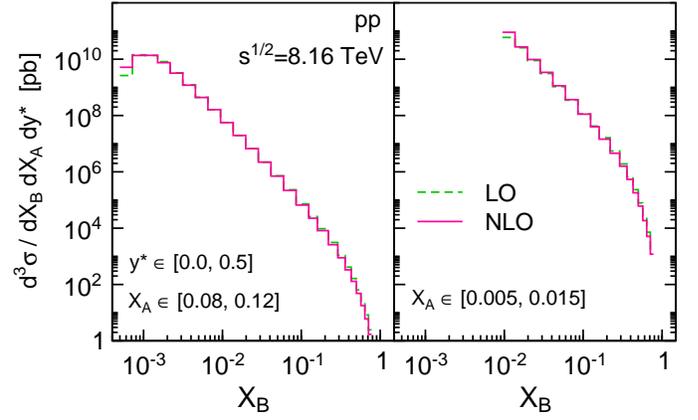}
\caption{Triple-differential cross section $d^3\sigma/dX_BdX_Ady^*$ versus $X_B$
in $pp$ collisions at $\sqrt{s}=8.16$~TeV.
Left and right panels correspond to two regions of $X_A$, $X_A\in [0.08, 0.12]$ and $X_A\in [0.005, 0.015]$, respectively.
The third variable $y^*$ is restricted with $y^*\in [0,0.5]$.
Theoretical results are calculated at LO~(dashed) and NLO~(solid).
Jets are found by anti-$k_T$ algorithm with cone size $R=0.7$ and with rapidity cut $|y|<2.5$.
Relative azimuthal angle of jet pair is restricted with $|\Delta\phi_{12}|>2\pi/3$.
Cuts imposed on transverse momenta of leading and sub-leading jets are $p_{T1}>30$~GeV and $p_{T2}>20$~GeV, respectively~\cite{CMS:2018jpl}.
}
\label{fig:XB_pp}
\end{figure}
Another choice of dijet variables to define a triple-differential cross section is $\{v_1,v_2,v_3\}=\{X_A,X_B,y^*\}$,
proposed by Ellis and Soper for the study of parton distributions in $pp(\bar{p})$ collisions~\cite{Ellis:1994dg},
with two new variables defined as
\bea
X_{A}\!=\!\sum_{n\in \textrm{dijet}}\!\frac{E_{Tn}}{\sqrt{s}}e^{+ y_n},~~~~
X_{B}\!=\!\sum_{n\in \textrm{dijet}}\!\frac{E_{Tn}}{\sqrt{s}}e^{- y_n},
\label{eq:XAXB}
\eea
which are similar as the expression of $x_{a(b)}$ in Eq.~(\ref{eq:inferx}).
Here the summation is performed over all the particles inside the dijet cones.
Apparently, the LO relation $X_{A(B)}=x_{a(b)}$ provides a direct connection to the initial momentum fractions.
At NLO, it still holds for the case that all the partons fall into the dijet cones, and
becomes $X_{A(B)}\leq x_{a(b)}$ only when there is one parton lying outside.
In Ref.~\cite{Ellis:1994dg}, it is shown that the NLO corrections on the cross section $d^3\sigma/dX_BdX_Ady^*$ can be small.

We first calculate the cross sections in $pp$ collisions at $\sqrt{s}=8.16$~TeV and show the results
as functions of $X_B$ in Fig.~\ref{fig:XB_pp}, where the left and right panels correspond to the regions $X_{A}\sim0.1$
and $X_{A}\sim0.01$, respectively. By restricting $X_A$, one actually controls the $x_a$ carried by the forward-going
initial parton~(the "probe" of nPDFs in $p$A). We find that, for $X_A\sim0.1$, dijet production covers a wide range of $X_B$~(from $10^{-3}$ to 0.8),
where the LO and NLO results are close to each other. We also check the dijet yields in each $\Delta X_B\Delta X_A\Delta y^*$ bin~(integrated cross section),
and find the yields here are similar as the those in Fig.~\ref{fig:ptavg_8_cms_pp}, and can be even larger for high-$x$ region.
By lowering $X_A$ to be around $0.01$, as shown in the right panel in Fig.~\ref{fig:XB_pp}, we find the dijet yields in large-$X_B$ regoin can
increase approximately by two orders of magnitude, indicating a higher statistical accuracy for
probing the large-$x$ parton distributions.
Since we have limited the jet transverse momenta as $p_{T1}>30$~GeV and $p_{T2}>20$~GeV~\cite{CMS:2018jpl},
the $X_B$ is unlikely to be very small for a lower $X_A$.
\begin{figure}[t]
\vspace{-0.6cm}\includegraphics[width=2.8in]{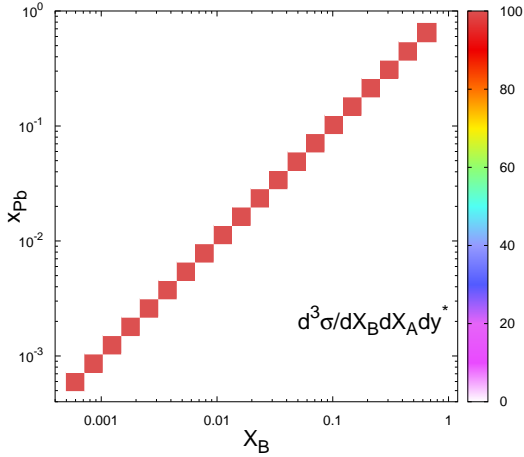}
\caption{Similar as Fig.~\ref{fig:ptavg_cr_lo_grid}, but for correlations between $X_B$ and $x_{\textrm{Pb}}$
in triple-differential cross section $d^3\sigma/dX_BdX_Ady^*$, corresponding to left panel of Fig.~\ref{fig:XB_pp}.
Bin sizes for $X_B$ and $x_{\textrm{Pb}}$ are set to be same.}
\label{fig:XB_cr_lo_grid}
\end{figure}

\begin{figure}[!h]
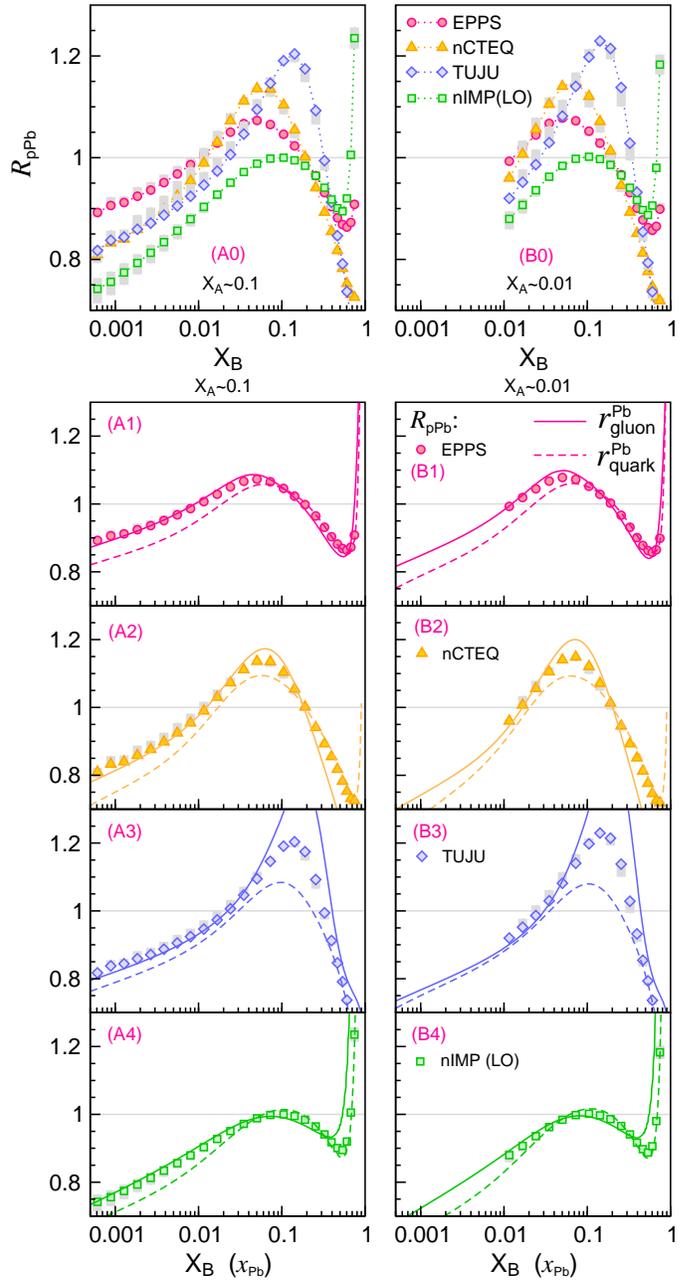

\hspace{-0.4cm}\includegraphics[width=3.5in]{XB_pp_rpa.eps}
\vspace{0.4cm}\hspace{0cm}
\includegraphics[width=3.3in]{XB_XA_rpa_rnpdf.eps}
\caption{Top panels (A0) and (B0): nuclear modification factors $R_{p\textrm{Pb}}$ for $X_A\sim0.1$ and $X_A\sim0.01$
corresponding to Fig.~\ref{fig:XB_pp}, predicted by four nPDF sets and plotted with symbols plus dotted curves.
Results of $R_{p\textrm{Pb}}$ in panels (A0) and (B0) are separately shown in (A1-4) and (B1-4) with symbols,
and compared with corresponding nPDF factors $r_i^{\textrm{Pb}}(x_{\textrm{Pb}}, Q^2)$ for gluon~(solid) and quark~(dashed) distributions.
One can note LO approximation $X_B=x_{\textrm{Pb}}$ in this comparison.
When plotting $r_i^{\textrm{Pb}}(x_{\textrm{Pb}}, Q^2)$, we have set $Q=\sqrt{X_AX_Bs}/[4\cosh(0.7y^*)]$ with $X_A=0.1~(0.01)$ and $y^*=0$.
Grey bands on $R_{p\textrm{Pb}}$ results correspond to variations with factorization/renormalization scales ($\mu_0/2, \mu_0, 2\mu_0$).}
\label{fig:XB_pp_rpa}
\end{figure}

In our calculations, the factorization/renormalization scale is taken to be $\mu_0\!=\!\sqrt{X_AX_Bs}/[4\cosh(0.7y^*)]$~\cite{Ellis:1994dg}.
Besides, the relative azimuthal angle of the two jets is restricted as $|\Delta\phi_{12}|\!>\!2\pi/3$~\cite{CMS:2018jpl}, through which
the jet pair is selected to be nearly back to back. This constraint is imposed to reduce the contributions of the
case that partons lie outside the dijet cones~(but become useless for the partons co-linearly emitted by initial partons).

Next, let us discuss for the $p$Pb collisions.
The correlations between $X_B$ and $x_{\textrm{Pb}}$ at LO are plotted in Fig.~\ref{fig:XB_cr_lo_grid}, and a perfect linear
correspondence can be observed as expected. Especially, it doesn't depend on the bin sizes of $\Delta X_A$ and $\Delta y^*$.
This high resolution may allow a more detailed kinematic scan of the nuclear PDF factors $r_i^{\textrm{Pb}}(x_{\textrm{Pb}}, Q^2)$.

The nuclear modification factors $R_{p\textrm{Pb}}$ on the triple-differential cross sections $d^3\sigma/dX_BdX_Ady^*$
are calculated with four nPDF sets~(EPPS, nCTEQ and TUJU at NLO and nIMP at LO), and shown in top panels (A0) and (B0)
of Fig.~\ref{fig:XB_pp_rpa} for $X_A\sim0.1$ and $X_A\sim0.01$, respectively.
A clear $X_B$ dependence is observed, and the results here are similar as the $r_i^{\textrm{Pb}}(x_{\textrm{Pb}}, Q^2)$
shown in Fig.~\ref{fig:ratio_nPDF}. With the LO approximation $X_B=x_{\textrm{Pb}}$ in mind, we can roughly identify the
shadowing, anti-shadowing, EMC, and Fermi motion effects in the observable $R_{p\textrm{Pb}}$.

The results in panels (A0) and (B0) with four nPDF sets are then separately shown in panels (A1-4) and (B1-4), and
are compared to their corresponding nPDF factors $r_i^{\textrm{Pb}}(x_{\textrm{Pb}}, Q^2)$ for gluon and quark distributions.
We see a very nice agreement between the $R_{p\textrm{Pb}}$ and $r_i^{\textrm{Pb}}$.
In particular, the $R_{p\textrm{Pb}}$ approaches to $r_{\textrm{gluon}}^{\textrm{Pb}}$ for gluon distribution
at small $X_B$~($\lesssim 0.01$), and is close to $r_{\textrm{quark}}^{\textrm{Pb}}$ for quark distribution
at large $X_B$~($\gtrsim 0.3$). This is reasonable, since the dijet production at small $X_B$ is dominated by the
nuclear gluon initiated processes, whereas at large $X_B$ the nuclear (valence)~quarks play a dominant role.
We also see that, in the intermediate $X_B$ region, $R_{p\textrm{Pb}}$ results from the interplay of $r_{\textrm{gluon}}^{\textrm{Pb}}$
and $r_{\textrm{quark}}^{\textrm{Pb}}$. In this sense, the $R_{p\textrm{Pb}}(X_B)$ on the observable level can serve as
an image of the nuclear modifications on parton distributions, i.e., the $r_i^{\textrm{Pb}}$ factors, owing to
the high resolution power for $x_{\textrm{Pb}}$ of the cross section $d^3\sigma/dX_BdX_Ady^*$.

On the other hand, we note that the physical scales will vary with the $X_B$ in this cross section, which means
the nuclear modifications at different values of $X_B$ are probed at different scales. For example, at LO one has
\bea
M_{{\!J\!J}}\!=\!\sqrt{X_AX_Bs},~~~
p_{T}\!=\!\sqrt{X_AX_Bs}/[2\cosh(y^*)].
\label{eq:XAXB_LO}
\eea
For given values of $X_A$ and $y^*$, both $M_{{\!J\!J}}$ and $p_T$ increase with increasing $X_B$.~(Note the choice of the
factorization scale $\mu_0$ in our calculation is a compromise between $M_{{\!J\!J}}/4$ and $p_T/2$~\cite{Ellis:1992en}).
As a matter of fact, if the values of both $M_{{\!J\!J}}$ and $p_T$ are fixed, the only remaining degree of freedom of
dijet kinematics at LO is the rapidity of dijet.
Next, we study the dijet rapidity related triple-differential cross section, in which the physical scales don't vary.

\subsection{Nuclear modifications on $d^3\sigma/dy_{{dijet}}dE_T^{\!J\!J}dp_{T,avg}$ and $d^3\sigma/dX_BdE_T^{\!J\!J}dp_{T,avg}$}
\label{sec:XB_ET}

\begin{figure}[t]
\vspace{-0.5cm}\includegraphics[width=2.8in]{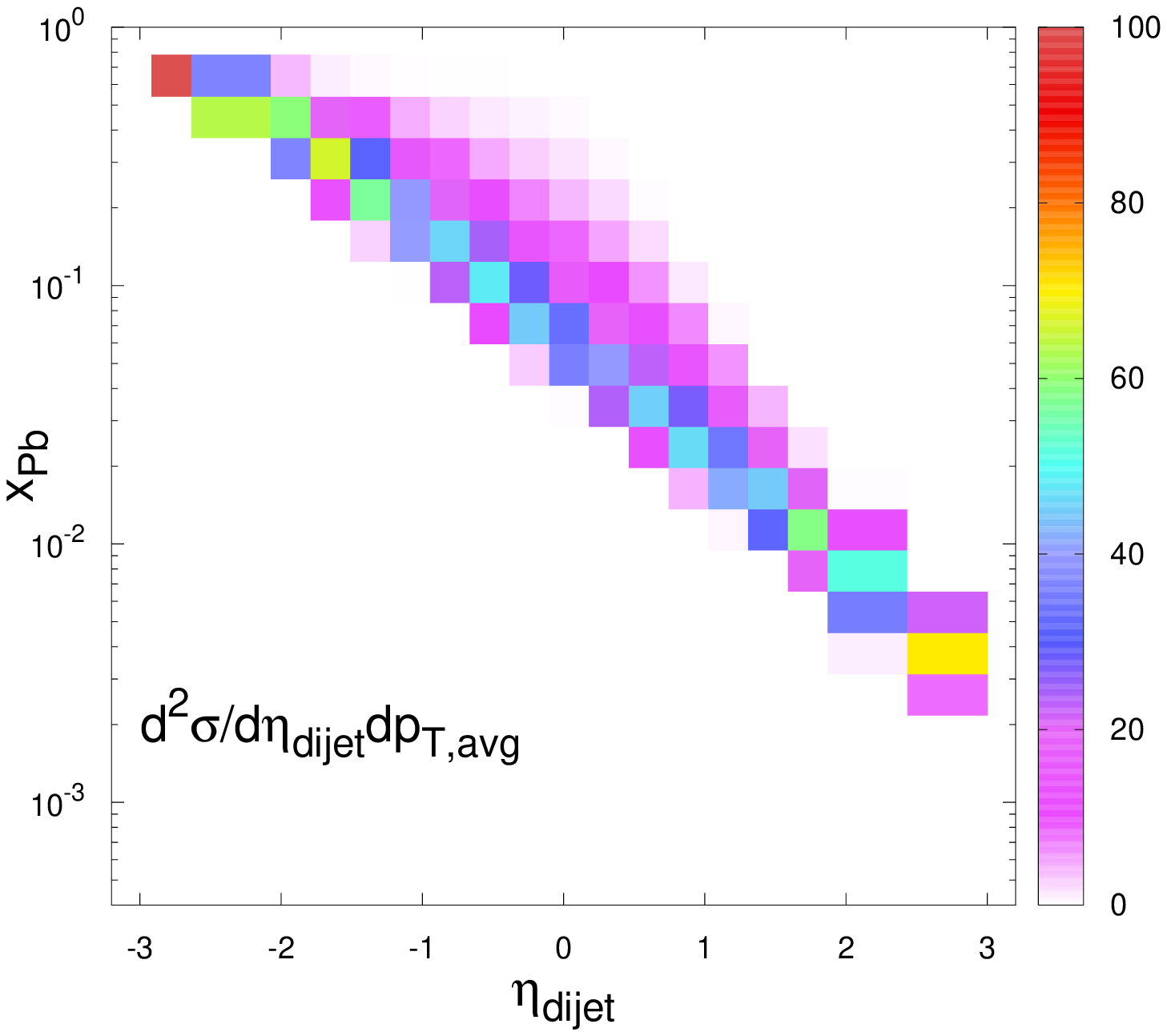}
\vspace{-0.5cm}\includegraphics[width=2.8in]{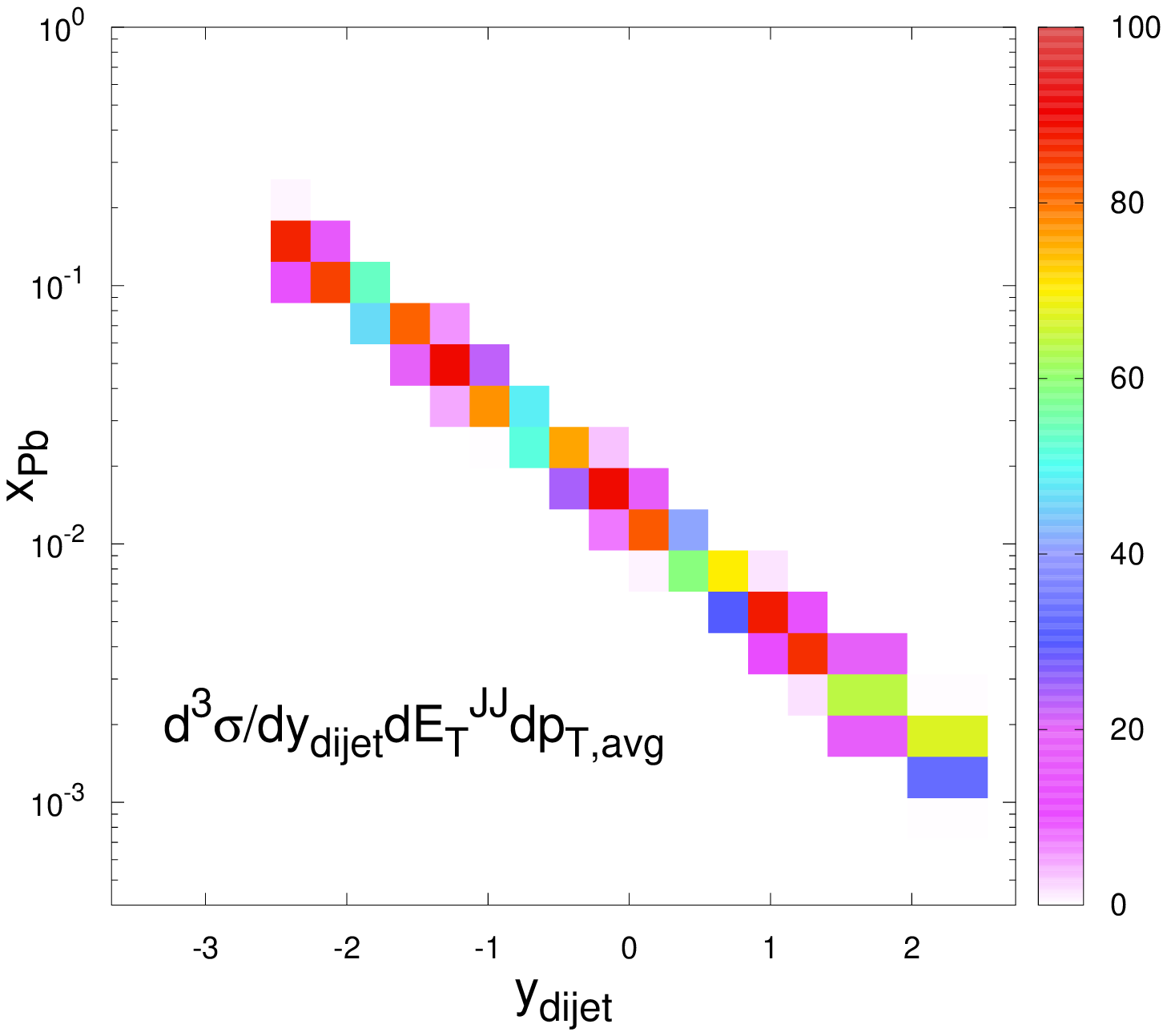}
\caption{Top panel: correlations between each $\eta_{\textrm{dijet}}$ bin and $x_{\textrm{Pb}}$ in double-differential
cross section $d^2\sigma/d\eta_{\textrm{dijet}}dp_{T,\textrm{avg}}$ corresponding to CMS measurement in
Figs.~\ref{fig:eta_cms_pp} and \ref{fig:eta_cms_rpa_sep}.
Bottom panel: correlations between each $y_{\textrm{dijet}}$ bin and $x_{\textrm{Pb}}$ in triple-differential
cross section $d^3\sigma/dy_{\textrm{dijet}}dE_T^{JJ}dp_{T,\textrm{avg}}$ corresponding to first panel in Fig.~\ref{fig:mT_pp}.
Bin sizes for $\eta_{\textrm{dijet}}$ and $y_{\textrm{dijet}}$ are set to be same. See also caption of
Fig.~\ref{fig:ptavg_cr_lo_grid} for more details of calculations.}
\label{fig:eta_cr_lo_grid}
\end{figure}

\begin{figure*}[t]
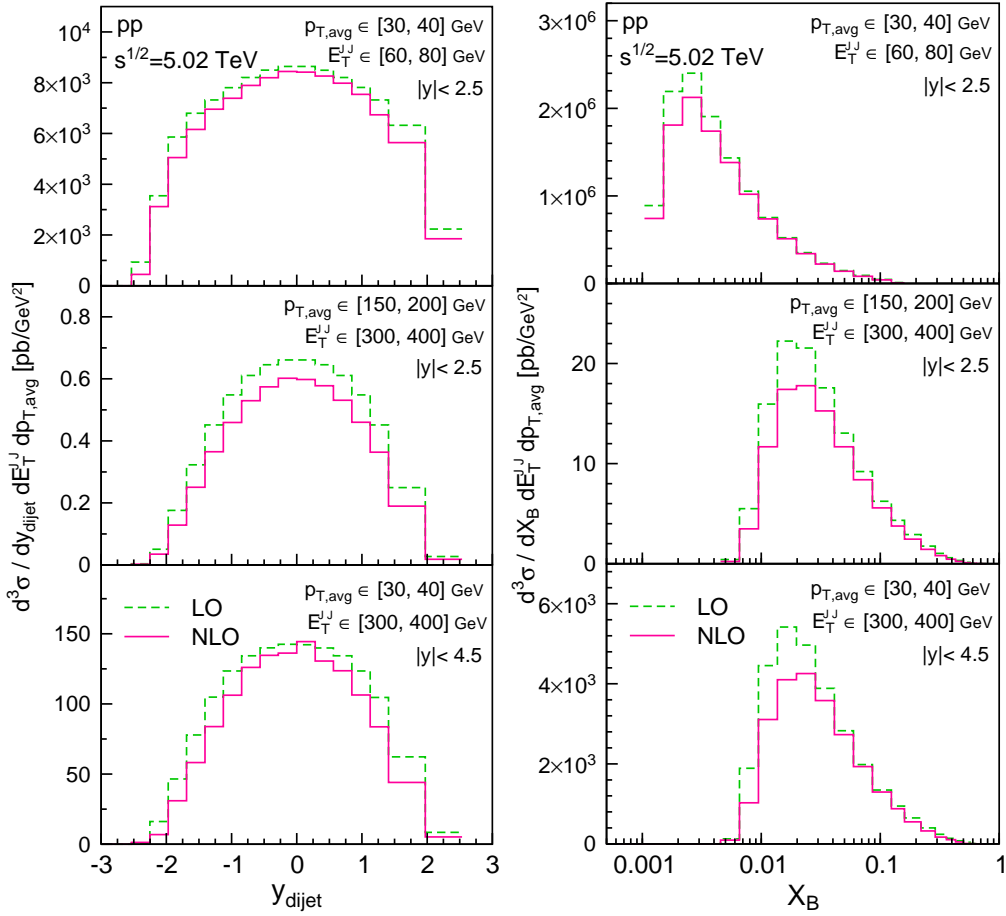

\hspace{-0.5cm}\includegraphics[width=2.6in]{mT_pp.eps}
\includegraphics[width=2.6in]{mT_XB_pp.eps}
\caption{Left panel: triple-differential dijet cross sections $d^3\sigma/dy_{\textrm{dijet}}dE_T^{\!J\!J}dp_{T,\textrm{avg}}$
in $pp$ collisions at $\sqrt{s}=5.02$~TeV calculated at LO~(dashed) and NLO~(solid). Results are calculated for
three $\{p_{T,\textrm{avg}},E_T^{{\!J\!J}}\}$ regions and shown in top, middle, and bottom plots,
respectively~(see annotations in each plot for details).
Right panel: similar as left one, but for $d^3\sigma/dX_BdE_T^{{\!J\!J}}dp_{T,\textrm{avg}}$.
Kinematic restrictions for left and right plots are set to be same.
%Compared to LO results, smaller cross sections are given by NLO calculations in three regions.
Jets are found by anti-$k_T$ algorithm with cone size $R=0.7$.
Relative azimuthal angle of jet pair is restricted with $|\Delta\phi_{12}|>2\pi/3$.
Cuts imposed on transverse momenta of leading and sub-leading jets are $p_{T1}>30$~GeV and $p_{T2}>20$~GeV, respectively~\cite{CMS:2018jpl}.
}
\label{fig:mT_pp}
\end{figure*}

A triple-differential cross section can also be defined with the rapidity of dijet.
Actually we have shown the dijet pseudo-rapidity distribution as measured by CMS in
Sec.~\ref{sec:pp}-\ref{sec:CNM}~(see Figs.~\ref{fig:eta_cms_pp} and \ref{fig:eta_cms_rpa_sep}).
As has been discussed, the measured distribution is related to the double-differential cross section
$d^2\sigma/d\eta_{\textrm{dijet}}dp_{T,\textrm{avg}}$, in which neither $x_{a(b)}$ nor $M_{{\!J\!J}}$ are fully controlled
at a given set of $\{\eta_{\textrm{dijet}},p_{T,\textrm{avg}}\}$.
It can be improved by measuring a triple-differential one with $V^{(3)}=\{\eta_{\textrm{dijet}},M_{{\!J\!J}},p_{T,\textrm{avg}}\}$.
Nevertheless, here we note the kinematic relation
\bea
X_B=\frac{1}{\sqrt{s}}E_T^{{\!J\!J}}\exp(-{y_{\textrm{dijet}}}),
\label{eq:XB_EY}
\eea
where $X_B$ is defined in Eq.~(\ref{eq:XAXB}), $E_T^{{\!J\!J}}$ and $y_{\textrm{dijet}}$ are
transverse energy and rapidity of dijet defined as
\bea
E_T^{{\!J\!J}}\!=\!\sqrt{M^2_{{\!J\!J}}\!+\!(\vec{p}_{T1}\!+\!\vec{p}_{T2})^2},~~
y_{\textrm{dijet}}\!=\!\frac{1}{2}\ln\frac{E^{\textrm{dijet}}\!+\!p_z^{\textrm{dijet}}}{E^{\textrm{dijet}}\!-\!p_z^{\textrm{dijet}}}.
\label{eq:EJJY}
\eea
Since Eq.~(\ref{eq:XB_EY}) holds for all orders in perturbative calculations, we defined an alternative dijet rapidity related
cross section binned with $V^{(3)}=\{y_{\textrm{dijet}},E_T^{{\!J\!J}},p_{T,\textrm{avg}}\}$, which has a more close connection to the variable $X_B$.
At LO, this choice is equivalent to the $V^{(3)}=\{\eta_{\textrm{dijet}},M_{{\!J\!J}},p_{T,\textrm{avg}}\}$, by noting
that $E_T^{{\!J\!J}}=M_{{\!J\!J}}$ and $y_{\textrm{dijet}}=\eta_{\textrm{dijet}}$. Differences between the two choices
only come from the high-order contributions.

To show the abilities of the observables to resolve $x_{\textrm{Pb}}$ in $p$Pb collisions,
in Fig.~\ref{fig:eta_cr_lo_grid}, we compare the correlations between $\eta_{\textrm{dijet}}$
and $x_{\textrm{Pb}}$ in the double-differential cross section with $V^{(2)}=\{\eta_{\textrm{dijet}},p_{T,\textrm{avg}}\}$,
to those between $y_{\textrm{dijet}}$ and $x_{\textrm{Pb}}$ in the newly defined triple-differential one with
$V^{(3)}=\{y_{\textrm{dijet}},E_T^{{\!J\!J}},p_{T,\textrm{avg}}\}$.
For the double-differential one, we observe a roughly linear correspondence between $\eta_{\textrm{dijet}}$ and $\log x_{\textrm{Pb}}$
resulting from $\eta_{\textrm{dijet}}=\frac{1}{2}(\ln x_a-\ln x_{\textrm{Pb}})$ at LO.
However, it is smeared due to the remaining degree of freedom, e.g. $x_a$ is not fixed.
In contrast, we see a clear linear correlation in the triple-differential one, indicating a better resolution of $x_{\textrm{Pb}}$.
Here the correlations between $y_{\textrm{dijet}}$ and $x_{\textrm{Pb}}$ are not exactly $100\%$,
because the bin sizes of them are not fully matched, e.g., the $\log x_{\textrm{Pb}}$ bins are equidistant whereas the $y_{\textrm{dijet}}$
are set to be the same as the $\eta_{\textrm{dijet}}$ in CMS measurement, and also because the bin sizes of
$\Delta E_T^{{\!J\!J}}$ and $\Delta p_{T,\textrm{avg}}$ are finite.

\begin{figure*}[t]
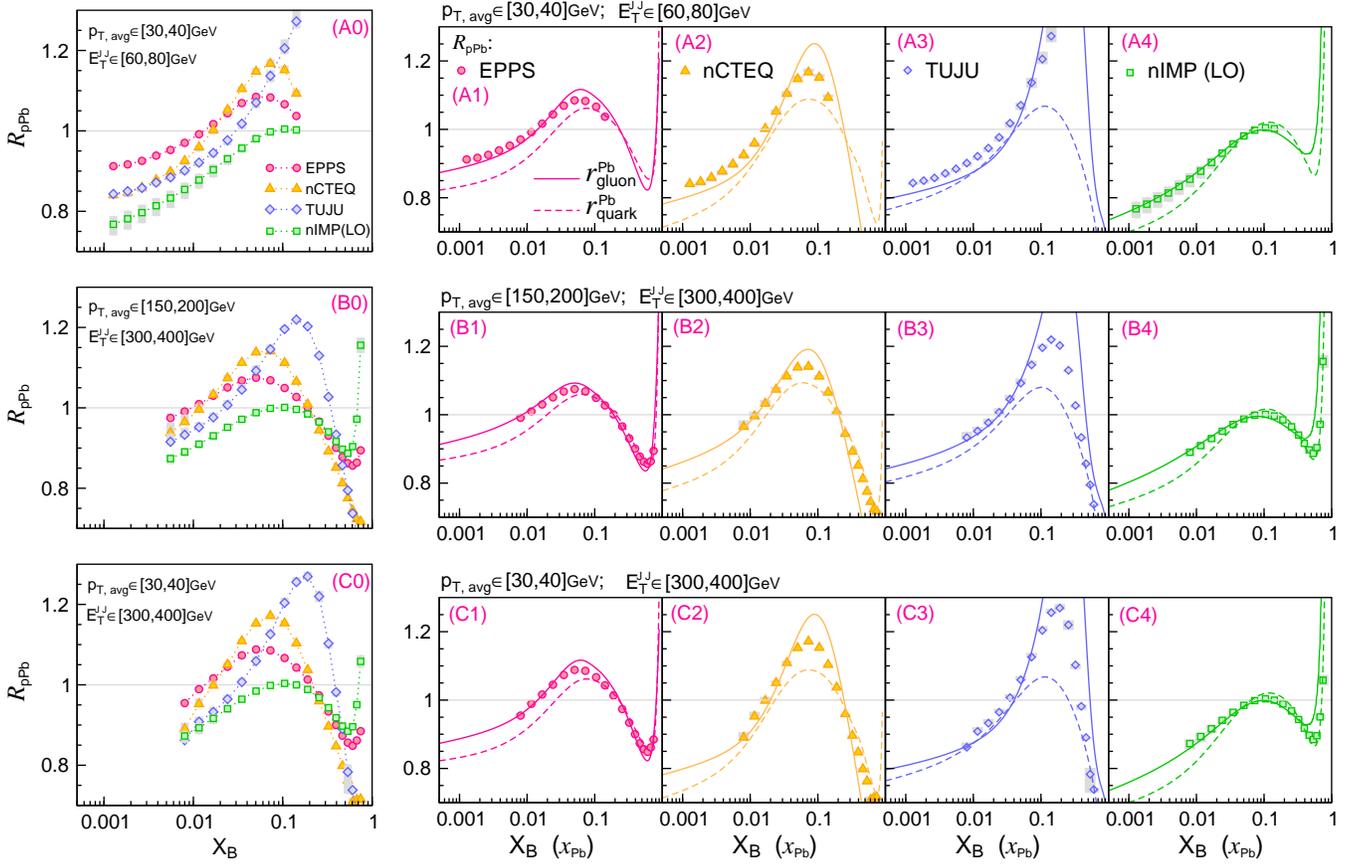

%   Requires \usepackage{graphicx}
  \hspace{-0.1cm}\includegraphics[width=1.95in]{mT_XB_pp_rpa.eps}
  \hspace{0.2cm}\includegraphics[width=4.9in]{mT_XB_rpa_rnpdf.eps}
  \caption{Left panels (A0), (B0) and (C0) show nuclear modification factors $R_{p\textrm{Pb}}$ of cross sections $d^3\sigma/dX_BdE_T^{{\!J\!J}}dp_{T,\textrm{avg}}$, corresponding to Fig.~\ref{fig:mT_pp},
  predicted by four nPDF sets and plotted with symbols plus dotted curves.
Results of $R_{p\textrm{Pb}}$ in panels (A0), (B0) and (C0) are separately shown in (A1-4), (B1-4) and (C1-4) with symbols,
and compared with corresponding nPDF factors $r_i^{\textrm{Pb}}(x_{\textrm{Pb}}, Q^2)$ for gluon~(solid) and quark~(dashed) distributions.
One can note LO approximation $X_B=x_{\textrm{Pb}}$ in this comparison.
When plotting $r_i^{\textrm{Pb}}(x_{\textrm{Pb}}, Q^2)$, we have set $Q=p_{T,\textrm{avg}}/2$~(15, 75, and 15~GeV for three regions).
Another choice $Q=E_T^{{\!J\!J}}/[4\cosh(0.7y^*)]$~(15, 75, and 30~GeV for three regions) is also
tested but not shown, with which a slightly better agreement between $R_{p\textrm{Pb}}$ and $r_i^{\textrm{Pb}}$ in third region is found.
Grey bands on $R_{p\textrm{Pb}}$ results correspond to variations with factorization/renormalization scales
($\mu_0/2, \mu_0, 2\mu_0$), which are seen to be small.
  }\label{fig:mT_XB_pp_rpa}
\end{figure*}

In the left-hand-side panel of Fig.~\ref{fig:mT_pp}, we show the cross sections
$d^3\sigma/dy_{\textrm{dijet}}dE_T^{{\!J\!J}}dp_{T,\textrm{avg}}$ versus $y_{\textrm{dijet}}$ in $pp$ collisions at $\sqrt{s}=5.02$~TeV,
calculated at LO and NLO for three kinematic regions distinguished with the $\{p_{T,\textrm{avg}},E_T^{{\!J\!J}}\}$ values.
Please note that, for two jets with equal rapidity, one has $E_T^{{\!J\!J}}=2p_{T,\textrm{avg}}$ at LO.
Thus the top and middle plots correspond to the case that $y_1\approx y_2$, on which a cut $|y_{1,2}|<2.5$ is imposed.
The bottom one corresponds to the case that two jets have a large rapidity difference~(
$|y_1-y_2|\sim4.5$ estimated with $E_T^{{\!J\!J}}/2p_{T,\textrm{avg}}=\cosh y^*$), thus a wider jet rapidity range $|y_{1,2}|<4.5$
is considered. Besides, since the middle and bottom plots share a same $E_T^{{\!J\!J}}$ bin, they actually probe in the same
$x_{\textrm{Pb}}$ range, but with different $p_{T,\textrm{avg}}$.
In Fig.~\ref{fig:mT_pp}, we can also see that the NLO calculations give somewhat smaller cross sections than the LO ones.
The factorzation/renormalization scale in the calculations is taken to be
$\mu_0=E_T^{{\!J\!J}}/[4\cosh(0.7y^*)]$~\cite{Ellis:1992en}, which is a compromise between $M_{{\!J\!J}}/4$ and $p_T/2$ in LO kinematics.

Utilizing the correspondence between $y_{\textrm{dijet}}$ and $X_B$ in Eq.~(\ref{eq:XB_EY}), one can easily transform the
dijet variable $V^{(3)}$ from $\{y_{\textrm{dijet}},E_T^{{\!J\!J}},p_{T,\textrm{avg}}\}$ to $\{X_B,E_T^{{\!J\!J}},p_{T,\textrm{avg}}\}$.
The transformed cross sections as functions of $X_B$, are plotted in the right-hand-side panel of Fig.~\ref{fig:mT_pp}.
Since the top plot is calculated for a lower $E_T^{{\!J\!J}}$, we can see that the dijets are produced in a smaller $X_B$ region,
compared to the middle and bottom ones. It is noteworthy that the LO correlations between $X_B$ and $x_{\textrm{Pb}}$ in $p$Pb collisions
should be the same as shown in Fig.~\ref{fig:XB_cr_lo_grid}, i.e., a one-to-one linear correspondence~(independent of the bin
sizes of $\Delta E_T^{{\!J\!J}}$ and $\Delta p_{T,\textrm{avg}}$).

The nuclear modification factors $R_{p\textrm{Pb}}$ of the cross sections $d^3\sigma/dX_BdE_T^{{\!J\!J}}dp_{T,\textrm{avg}}$,
corresponding to the results in Fig.~\ref{fig:mT_pp}, are calculated at NLO with several nPDF sets~(except for nIMP at LO), and
shown in panels (A0), (B0) and (C0) of Fig.~\ref{fig:mT_XB_pp_rpa}. The results are very similar as those in Fig.~\ref{fig:XB_pp_rpa}.
These $R_{p\textrm{Pb}}$ predicted by four nPDF sets are then separately shown in panels (A1-4), (B1-4) and (C1-4), and are
compared to their corresponding nPDF factors $r_i^{\textrm{Pb}}(x_{\textrm{Pb}}, Q^2)$ for gluon and quark distributions.
By noting the LO approximation $X_B=x_{\textrm{Pb}}$,
we see again that the $R_{p\textrm{Pb}}(X_B)$ results provide overall nice images of the $r_i^{\textrm{Pb}}(x_{\textrm{Pb}})$.
With the increasing $X_B$, we can observe the transition from gluon-dominated to quark-dominated regions.
In particular, considering that the physical scales are well controlled, we have plotted the $r_i^{\textrm{Pb}}(x_{\textrm{Pb}},Q^2)$
in Fig.~\ref{fig:mT_XB_pp_rpa} at fixed values of $Q^2$, not like the $Q^2$ increasing with $X_B$ in Fig.~\ref{fig:XB_pp_rpa}.

In panels (A1-3) of Fig.~\ref{fig:mT_XB_pp_rpa}, we can see that the $R_{p\textrm{Pb}}$ at small values of
$X_B$~(shadowing region) are slightly higher than the
$r_{\textrm{gluon}}^{\textrm{Pb}}$ for gluon. This deviation may be partly attributed to the NLO process that
a parton is co-linearly emitted by the initial nuclear parton. In this case, $X_B<x_{\textrm{Pb}}$,
which means the nuclear correction from a larger value of $x_{\textrm{Pb}}$, e.g., in anti-shadowing region, may enter the $R_{p\textrm{Pb}}$.
As is expected, this effect is not observed in the nIMP results at LO shown in panel (A4).
In panels (B1-4) and (C1-4), this effect is seen to be weaker.
Please note that the values of the physical scales, $E_T^{{\!J\!J}}~(M_{{\!J\!J}})$ and $p_{T,\textrm{avg}}$,
in these panels are larger than those in (A1-4). At a higher probing scale, the nuclear partons tend to carry smaller momentum fraction $x$
and the contributions from the initial-state co-linear emission could be suppressed.

Since the $R_{p\textrm{Pb}}(X_B)$ well reflects the $r_i^{\textrm{Pb}}(x_{\textrm{Pb}},Q^2)$ at a certain value of $Q^2$,
it will be an interesting question that, whether the $R_{p\textrm{Pb}}(X_B)$ measured at different $p_{T,\textrm{avg}}$ values
can shed light on the variation of $r_i^{\textrm{Pb}}(x_{\textrm{Pb}},Q^2)$ with $Q^2$?

\begin{figure}[t]
\hspace{-0.5cm}\includegraphics[width=3.3in]{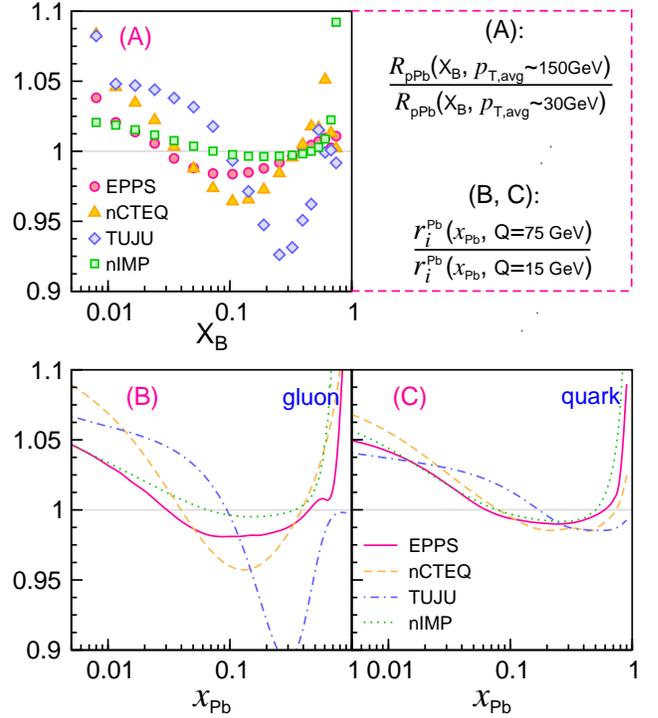}
\caption{Panel (A): ratio of $R_{p\textrm{Pb}}(X_B)$ in two regions of $p_{T,\textrm{avg}}$, i.e.,
$p_{T,\textrm{avg}}\in[150,200]$~GeV and $[30,40]$~GeV,
corresponding to panels (B0) and (C0) in Fig.~\ref{fig:mT_XB_pp_rpa}.
Panels (B) and (C): ratios of nPDFs factors $r_i^{\textrm{Pb}}(x_{\textrm{Pb}},Q^2)$
at two values of $Q$~($75$ and $15$~GeV, taken as $Q\sim p_{T,\textrm{avg}}/2$),
for gluon~(B) and quark~(C) distributions, respectively.}
\label{fig:mT23_XB_rsc}
\end{figure}

A straightforward test can be made by utilizing the results in panels (B0) and (C0) of Fig.~\ref{fig:mT_XB_pp_rpa},
which are in the same $X_B$ range but correspond to two different $p_{T,\textrm{avg}}$ regions, respectively.
The ratios of the $R_{p\textrm{Pb}}$ in the higher $p_{T,\textrm{avg}}$ region~($\sim150$~GeV) to those at the
lower $p_{T,\textrm{avg}}$~($\sim30$~GeV) are calculated and shown in panel (A) of Fig.\ref{fig:mT23_XB_rsc}.
We find the $R_{p\textrm{Pb}}$ given by four nPDF sets exhibit different variations with $p_{T,\textrm{avg}}$.
To further understand these distinctions, we plot in panels (B) and (C) the ratios of $r_i^{\textrm{Pb}}(x_{\textrm{Pb}},Q^2)$ at $Q=75$~GeV
to those at $Q=15$~GeV~(simply taken to be $p_{T,\textrm{avg}}/2$, neglecting the impact from $E_T^{{\!J\!J}}$)
for gluon and quark distributions, respectively.
We observe that, for quark distribution, the variations of the $r_{\textrm{quark}}^{\textrm{Pb}}$
from the four nPDF sets have small distinctions.
In contrast, for gluon distribution, significant differences among the results can be observed,
which are similar as the differences seen in the variations of $R_{p\textrm{Pb}}$ with $p_{T,\textrm{avg}}$.

These results are somehow in accordance with the theoretical expectation
from the collinear factorization of perturbative QCD~\cite{Collins:1989gx}.
Here, the dependence of $r_i^{\textrm{Pb}}(x_{\textrm{Pb}},Q^2)$ on $Q^2$ results from the
separate QCD evolutions of the nuclear PDFs~(i.e.~EPPS, nCTEQ, TUJU, and nIMP) and free-nucleon PDFs governed by the
Dokshitzer-Gribov-Lipatov-Altarelli-Parisi~(DGLAP) equations. It reflects the fact
that at different resolution scales the nuclear modifications on partonic structures can be different
in general~(see Fig.~\ref{fig:rnpdfQ} in Appendix \ref{append:rscale}).
The results in Fig.~\ref{fig:mT23_XB_rsc} indicate that the variations of the measured $R_{p\textrm{Pb}}(X_B)$,
at different kinematic energy scales of the hard scattering~(related to $p_{T,\textrm{avg}}$),
is correlated to the underlying energy-scale dependence of nuclear modifications to some extent.
In the context of QCD factorization~\cite{Collins:1989gx}, this is related to the feature that the
renormalization group equation of the hard scattering cross section $H_{ab}$, which involves the
kinematic energy scale, shares the same evolution kernels as in the DGLAP equation for PDFs.
\begin{figure*}[t]
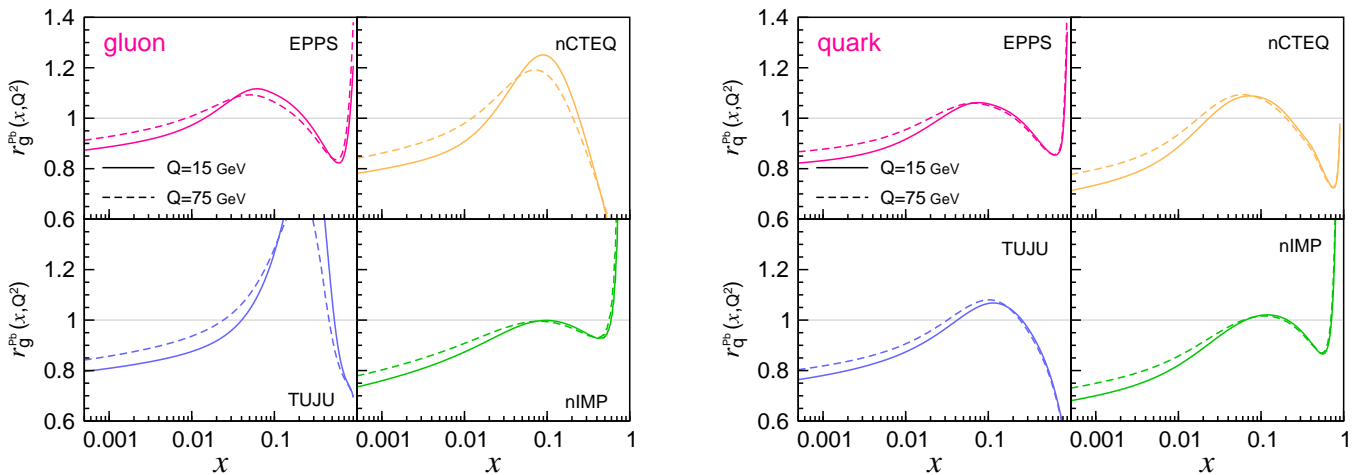

\hspace{-0.1cm}\includegraphics[width=3.3in]{npdf_sc_g}
\hspace{1.0cm}\includegraphics[width=3.3in]{npdf_sc_q}
\caption{Nuclear correction factors $r^{\textrm{Pb}}_i(x,Q^2)$ for gluon~(left half) and quark~(right half) distributions
at $Q=15$ and $75$~GeV from four nPDF sets, EPPS, nCTEQ, TUJU and nIMP.}
\label{fig:rnpdfQ}
\end{figure*}

Relevant to this feature, we have also tested other choices of factorization/renormalization
scale $\mu$ in the calculations of $R_{p\textrm{Pb}}(X_B)$, and found that at NLO their
variations with $p_{T,\textrm{avg}}$ are rather insensitive to the choice of $\mu$.
This is expected, since in finite-order approximation the $\mu$-dependence of the factorized
cross section is on the magnitude of the (uncalculated) higher-order contributions.
For the $R_{p\textrm{Pb}}$ defined as a ratio, the $\mu$-dependence is further suppressed.
In addition, when comparing $R_{p\textrm{Pb}}$ to $r_i^{\textrm{Pb}}(x_{\textrm{Pb}},Q^2)$
in this and previous subsections, we have approximated the values of
$Q^2$ in $r_i^{\textrm{Pb}}$ on the order of the kinematic energy scales
of the hard process for a reasonable comparison.
In calculations for cross sections, $\mu$ is taken on the same order of energy scale
to ensure the efficacy of the perturbation theory.

All in all, the results in this subsection demonstrate the advantage of the triple-differential measurement
binned with $\{X_B,E_T^{{\!J\!J}},p_{T,\textrm{avg}}\}$ for a detailed kinematic scan of the
underlying nuclear modification $r_i^{\textrm{Pb}}(x_{\textrm{Pb}},Q^2)$,
including both the $x_{\textrm{Pb}}$ and $Q^2$ dependence.

~

~

%%%%%%%%%%%%%%%%%%%%%%%%%%%%%%%%%%%%%%%%%%%%%%%
\section{Summary and Discussion}
\label{sec:summary}

The deviations of the nuclear parton distributions from those of a free nucleon, arising from the
additional non-perturbative dynamics that bind the nucleons together, can be generally quantified as the
correction factors $r_i^{A}(x,Q^2)={f^{A,p}_i(x,Q^2)}/{f^{p}_i(x,Q^2)}$ as functions of
the nuclear mass number $A$, parton flavor $i$, momentum fraction $x$ and resolution scale $Q^2$.
An irreplaceable approach to accessing the corrections is to extract $r_i^{A}(x,Q^2)$ from the experimental
measurements, relying on the factorization in perturbative QCD.
In a realistic observable~(e.g., a differential cross section), nuclear effects from different $x$ regions,
at different probing scales, and for various parton flavors usually mix together.
However, a properly defined observable that can faithfully resolve both $x$ and $Q^2$ at LO level, may
provide a more effective and detailed kinematic scan of the $x$ and $Q^2$ dependence of $r_i^{A}(x,Q^2)$.

In this work, we focus on the dijet production in $p$Pb collisions at the LHC as a probe of the
nuclear quark and gluon distributions. In order to well resolve the momentum fractions of initial-state partons
as well as the probing scale, we study several types of triple-differential cross sections in $pp$ and $p$Pb collisions,
including those binned with dijet variables $V^{(3)}=\{p_{T,\textrm{avg}},y_b,y^*\}$, $\{X_B,X_A,y^*\}$,
and $\{X_B,E_T^{{\!J\!J}},p_{T,\textrm{avg}}\}$. These cross sections are calculated to NLO in perturbative QCD.
Four sets of nPDFs, EPPS16, nCTEQ15, TUJU19, and nIMParton16 are employed in the calculations for $p$Pb.
The observable nuclear modification factors $R_{p\textrm{Pb}}$ of the triple-differential cross sections,
especially those as functions of $X_B$, are found to provide a nice image of the nPDF correction factors
$r_i^{A}(x,Q^2)$ from small to large values of $x$.
Based on this, the differences among the $R_{p\textrm{Pb}}$ predicted by various nPDFs can be well interpreted.
In particular, with $V^{(3)}=\{X_B,E_T^{{\!J\!J}},p_{T,\textrm{avg}}\}$, the variation of $R_{p\textrm{Pb}}$
with $p_{T,\textrm{avg}}$ can even reflect the subtle scale variation of $r_i^{A}(x,Q^2)$.

In future, the measurements of these $R_{p\textrm{Pb}}$ of triple-differential dijet cross sections
at the LHC with a high precision are expected to provide significant and multi-dimensional constraints
on various parametrization sets of nPDFs.
More importantly, the measurements will provide an effective way to help confirm various nuclear effects, e.g.,
shadowing, anti-shadowing, EMC, and Fermi motion in different regions of $x$ and their variations with probing scale $Q^2$.
It is also noteworthy that, the studied triple-differential cross sections can be straightforward generalized
and applied to the study of other processes, such as vector-boson-tagged jet production~\cite{Ma:2018tjv,Zhang:2018urd,
Zhang:2021oki} and heavy-quark dijet production~\cite{Dai:2018mhw}.
We also hope the more accurate description of the initial-state cold nuclear matter effects provides
a baseline to better understand the final-state jet quenching phenomena in relativistic heavy-ion collisions.

%%%%%%%%%%%%%%%%%%%%%%%%%%%%%%%%%%%%%%%%%%%%%%%
\begin{acknowledgements}
The authors would like to thank E.~Wang, H.-Z.~Zhang, H.~Xing, Y.-C.~He, W.~Dai, S.-Y.~Chen, G.-Y.~Ma,
S.-L.~Zhang, S.~Wang and Y.~Luo for helpful discussions.
This research was supported in part by Guangdong Major Project of Basic and Applied Basic Research
No.~2020B0301030008, by Natural Science Foundation of China~(NSFC) under Project No.~11935007,
and by Science and Technology Program of Guangzhou~(No.~2019050001).
P.R. is supported by China Postdoctoral Science Foundation under Project No.~2019M652929,
and the MOE Key Laboratory of Quark and Lepton Physics~(CCNU) under Project No.~QLPL201802.

%P.R. is supported by China Postdoctoral Science Foundation under project No. 2019M652929, Z.K. is supported by the National Science Foundation in US under Grant No. PHY-1720486, E.W., H.X. and B.Z. are supported by NSFC of China under Project No. 11435004.
\end{acknowledgements}

\begin{appendix}
\section{Variation of $r^{Pb}_i(x,Q^2)$ with $Q^2$}
\label{append:rscale}
Since both the free-nucleon and nuclear PDFs evolve with the resolution scale $Q^2$, the nuclear correction
factor $r^{A}_i(x,Q^2)$, i.e., the ratio of them, generally depends on $Q^2$. To visualize the variations of
$r^{A}_i(x,Q^2)$ with $Q^2$, we supplement in Fig.~\ref{fig:rnpdfQ} the $r^{\textrm{Pb}}_i(x,Q^2)$
for gluon and quark distributions at $Q=15$ and $75$~GeV, from four nPDF sets used in this work.

\label{section:appendix}

\end{appendix}

~

~

~

%%%%%%%%%%%%%%%%%%%%%%%%%%%%%%%%%%%%%%%%%%%%%%%%%%%%%%%%%%%%

 %%%%%%%%%%%%%%
%\bibliographystyle{h-physrev5}
%\bibliography{biblio}

\end{document}